\pdfoutput=1
\documentclass[times,preprint]{elsarticle}
\usepackage[a4paper,
            top=2.1cm,
            bottom=2.1cm,
            left=1.75cm,
            right=1.75cm,
            columnsep=18pt]{geometry}
\usepackage{multicol}     
\usepackage{framed,multirow}
\usepackage{amssymb}
\usepackage{latexsym}
\usepackage{url}
\usepackage{xcolor}
\usepackage{hyperref}
\usepackage{graphicx}
\usepackage{amsmath}
\usepackage{algorithmic}
\usepackage{array}
\usepackage{booktabs}
\usepackage{rotating}
\usepackage{float}
\usepackage{pbox}
\usepackage{rotating}
\usepackage{cleveref}

\usepackage{caption}

\Crefname{figure}{Fig.}{Figs.}
\Crefname{table}{Table.}{Tables.}
\Crefname{equation}{Equation.}{Equations.}
\definecolor{newcolor}{rgb}{.8,.349,.1}

\begin{document}
\begin{frontmatter}
\title{Extreme Cardiac MRI Analysis under Respiratory Motion: Results of the CMRxMotion Challenge}%
\author[1,2]{Kang Wang\fnref{fn1}}
\author[3]{Chen Qin\fnref{fn1}}
\author[4]{Zhang Shi\fnref{fn1}}
\fntext[fn1]{These authors contributed equally.}
\author[1,2]{Haoran Wang}
\author[1,2]{Xiwen Zhang}
\author[5,6]{Chen Chen}
\author[5,7]{Cheng Ouyang}
\author[5]{Chengliang Dai}
\author[7]{Yuanhan Mo} 
\author[4]{Chenchen Dai}
\author[8]{Xutong Kuang}
\author[9]{Ruizhe Li}
\author[9]{Xin Chen}
\author[10]{Xiuzheng Yue}
\author[10]{Song Tian}
\author[11,12]{Alejandro Mora-Rubio} 
\author[11]{Kumaradevan Punithakumar}
\author[13]{Shizhan Gong}
\author[13]{Qi Dou}
\author[14]{Sina Amirrajab}
\author[15]{Yasmina Al Khalil} 
\author[15]{Cian M. Scannell}
\author[16]{Lexiaozi Fan}
\author[17]{Huili Yang} 
\author[18]{Xiaowu Sun}
\author[18]{Rob van der Geest}
\author[19]{Tewodros Weldebirhan Arega}
\author[20]{Fabrice Meriaudeau} 
\author[21]{Caner Özer}
\author[22]{Amin Ranem} 
\author[22]{John Kalkhof} 
\author[21]{İlkay Öksüz} 
\author[22]{Anirban Mukhopadhyay}
\author[23]{Abdul Qayyum}
\author[24]{Moona Mazher}
\author[23]{Steven A Niederer}
\author[25]{Carles Garcia-Cabrera}
\author[26]{Eric Arazo} 
\author[27]{Michal K. Grzeszczyk}
\author[28]{Szymon Płotka}
\author[29]{Wanqin Ma}
\author[29]{Xiaomeng Li}
\author[30]{Rongjun Ge} 
\author[31]{Yongqing Kou}
\author[1,2,32,33]{Xinrong Chen}
\author[34]{He Wang}

\author[4]{Chengyan Wang\corref{cor1}}
\author[5,35,36]{Wenjia Bai\corref{cor1}}
\author[1,2]{Shuo Wang\corref{cor1}}
\cortext[cor1]{Corresponding authors: Shuo Wang (shuowang@fudan.edu.cn), Chengyan Wang (wangcy@fudan.edu.cn), Wenjia Bai (w.bai@imperial.ac.uk).}

\address[1]{Digital Medical Research Center, School of Basic Medical Sciences, Fudan University, Shanghai, Shanghai 200032, China}
\address[2]{Shanghai Key Laboratory of MICCAI, Fudan University, Shanghai, Shanghai 200032, China}
\address[3]{Department of Electrical and Electronic Engineering \& I-X, Imperial College London, London, London SW7 2AZ, United Kingdom}
\address[4]{Department of Radiology, Zhongshan Hospital Affiliated to Fudan University, Shanghai, Shanghai 200032, China}
\address[5]{Department of Computing, Imperial College London, London, London SW7 2AZ, United Kingdom}
\address[6]{School of Computer Science, University of Sheffield, Sheffield, S1 4DP, United Kingdom}
\address[7]{Department of Engineering Science, University of Oxford, Oxford, OX2 0ES, United Kingdom}
\address[8]{Shanghai Pudong Hospital and Human Phenome Institute, Fudan University, Shanghai, 201203, China}
\address[9]{School of Computer Science, University of Nottingham, Nottingham, NG8 1BB, United Kingdom}
\address[10]{Clinical \& Technical Support, Philips Healthcare, Beijing, Beijing 100600, China}
\address[11]{Radiology and Diagnostic Imaging, University of Alberta, Edmonton, AB T6G 1K4, Canada}
\address[12]{Departamento de Electrónica y Automatización, Universidad Autónoma de Manizales, Manizales 170001, Caldas, Colombia}
\address[13]{Department of Computer Science and Engineering, The Chinese University of Hong Kong, Hong Kong, Hong Kong 000000, China}
\address[14]{The D-Lab, Department of Precision Medicine, GROW - Research Institute for Oncology and Reproduction, Maastricht University, 6220 MD Maastricht, The Netherlands}
\address[15]{Department of Biomedical Engineering, Eindhoven University of Technology, Eindhoven 5612 AZ, The Netherlands}
\address[16]{Department of Radiology, Northwestern University, 737 N. Michigan Ave, Suite 1600, Chicago 60611, United States}
\address[17]{United Imaging Research, 393 Middle Huaxia Road, Pudong, Shanghai 201210, China}
\address[18]{Division of Image Processing, Department of Radiology, Leiden University Medical Center, PO Box 9600, Leiden 2300 RC, The Netherlands}
\address[19]{Université Bourgogne Europe, ImViA UR 7535, 21000 Dijon, France}
\address[20]{Université Bourgogne Europe, CNRS, ICMUB UMR 6302, 21000 Dijon, France}
\address[21]{Istanbul Technical University, Maslak, 34467, İstanbul, Türkiye}
\address[22]{Computer Science, Technical University of Darmstadt, Karolinenpl. 5, 64289 Darmstadt, Germany}
\address[23]{National Heart and Lung Institute, Faculty of Medicine, Imperial College London, Guy Scadding Building, Cale Street, London, SW3 6LY,United Kingdom}
\address[24]{Hawkes Institute, Department of Computer Science, University College London, 66-72 Gower St, London, United Kingdom}
\address[25]{School of Medicine, University College Dublin, Belfield, Dublin, D04 V1W8, Ireland}
\address[26]{CeADAR: Ireland's Centre for AI, University College Dublin, Belfield, Dublin, D04 V1W8, Ireland}
\address[27]{Sano Centre for Computational Medicine, Czarnowiejska 36, 30-054, Krakow, Poland}
\address[28]{Faculty of Mathematics and Computer Science, Jagiellonian University, S. Łojasiewicza 6, Krakow, Poland}
\address[29]{Department of Electronic and Computer Engineering, The Hong Kong University of Science and Technology, Clear Water Bay, Hong Kong 000, China}
\address[30]{School of Instrument Science and Engineering, Southeast University, Nanjing, Nanjing 210096, China}
\address[31]{College of Artificial Intelligence, Nanjing University of Aeronautics and Astronautics, Nanjing, Nanjing 211106, China}
\address[32]{Academy for Engineering and Technology, Fudan University, Shanghai, Shanghai 200433, China}
\address[33]{College of Biomedical Engineering, Fudan University, Shanghai, Shanghai 200433, China}
\address[34]{Institute of Science and Technology for Brain-inspired Intelligence, Fudan University, Shanghai, Shanghai 200433, China}
\address[35]{Department of Brain Sciences, Imperial College London, London, London SW7 2AZ, United Kingdom}
\address[36]{Data Science Institute, Imperial College London, London, London SW7 2AZ, United Kingdom}

\begin{abstract}
Deep learning models have achieved state-of-the-art performance in automated Cardiac Magnetic Resonance (CMR) analysis. However, the efficacy of these models is highly dependent on the availability of high-quality, artifact-free images. In clinical practice, CMR acquisitions are frequently degraded by respiratory motion, yet the robustness of deep learning models against such artifacts remains an underexplored problem.
To promote research in this domain, we organized the \textit{MICCAI CMRxMotion} challenge. We curated and publicly released a dataset of 320 CMR cine series from 40 healthy volunteers who performed specific breathing protocols to induce a controlled spectrum of motion artifacts. The challenge comprised two tasks: 1) automated image quality assessment to classify images based on motion severity, and 2) robust myocardial segmentation in the presence of motion artifacts. A total of 22 algorithms were submitted and evaluated on the two designated tasks. This paper presents a comprehensive overview of the challenge design and dataset, reports the evaluation results for the top-performing methods, and further investigates the impact of motion artifacts on five clinically relevant biomarkers. All resources and code are publicly available at: \href{https://github.com/CMRxMotion}{https://github.com/CMRxMotion}.
\end{abstract}


\end{frontmatter}

\begin{multicols}{2}


\section{Introduction}
Cardiac magnetic resonance (CMR) imaging is the gold-standard modality for evaluating cardiac structure and function~\citep{schulz2020standardized}. In terms of image analysis, deep learning approaches have achieved remarkable performance in automated CMR image segmentation~\citep{chen2020deep,bai2020population}. However, the generalizability of these segmentation models is still challenged by inconsistent imaging environments (e.g., different vendors or protocols)~\citep{9458279}, population shifts (normal vs. pathological cases)~\citep{ACDC} and unexpected human behaviors (e.g., body movements). To build a reliable segmentation model, it is useful to investigate when and how the model fails~\citep{wang2020deep} by exposing a trained segmentation model to extreme cases in a stress test~\citep{doi:10.1148/ryai.2021210097}, such as altered image quality caused by motion artifacts that may occur during clinical practice. To date, most CMR image segmentation challenges have focused on vendor variability~\citep{9458279}. and anatomical structure variations~\citep{ACDC, 10103611}, while the implications of human behaviors are less explored. For CMR image analysis, respiration motion is one of the major human behaviors that influence image quality~\citep{ferreira2013cardiovascular}. Patients may not be able to follow breath-hold instructions well, particularly those with heart failure or pediatric patients. Poor breath-hold behaviors result in degraded image quality and inaccurate analysis of cardiac structures~\citep{wang2021joint}. To establish a public benchmark dataset for assessing the effects of respiratory motion on CMR quality and examining the robustness of segmentation models~\citep{paschali2018generalizability}, we organized the extreme CMR image analysis challenge under respiratory motion challenge (\textit{CMRxMotion}) at \textit{MICCAI 2022} \footnote{\href{http://cmr.miccai.cloud}{http://cmr.miccai.cloud}}. 

\begin{table*}[hbtp]
  \centering
  \caption{Precedent related challenges for cardiac MR image segmentation. CMR: Cardiac Magnetic Resonance, SSFP: steady-state free precession sequences, bSSFP: balanced steady-state free precession sequences, LGE: Late gadolinium enhancement, MI: myocardial infarction, MH: myocardial hypertrophy, NHP: normal healthy subject, DCM: dilated cardiomyopathy, HCM: Hypertrophic Cardiomyopathy, CAM: Congenital Arrhythmogenesis, ARR: Arrhythmogenic cardiomyopathy, HF: Heart Failure, TF: Tetralogy of Fallot, AF: Atrial Fibrilation, IC: Interatrial Comunication, TR: Tricuspidal Regurgitation, ARV: abnormal right ventricle, LA: Left atrium, RA: Right atrium, LV: left ventrical, MYO: left ventrical mycardiomal, RV: right venticle, PV: Pulmonary vein, AO: Ascending aorta, SAX: short axis, LAX: long axis, N/A: Not reported, $\dagger$: The data reported only provided the number of MR images, without additional details.}
  \resizebox{\linewidth}{!}{
    \begin{tabular}{lp{14.335em}llp{6.375em}llllp{6.915em}p{11.625em}c}
      \toprule
      \multicolumn{1}{l}{\multirow{2}[4]{*}{Challenge}} & \multirow{2}[4]{*}{Challenge Name} & \multicolumn{1}{l}{\multirow{2}[4]{*}{Year}} & \multicolumn{1}{l}{\multirow{2}[4]{*}{Reference}} & \multicolumn{4}{>{\centering}p{20em}}{Data Information and Data Split} & \multicolumn{1}{c}{\multirow{2}[4]{*}{Segmentation Target}} & \multicolumn{1}{c}{\multirow{2}[4]{*}{Pathologies}} & \multicolumn{1}{c}{\multirow{2}[4]{*}{Scancer}} & \multicolumn{1}{c}{\multirow{2}[4]{*}{Held in Conjunction }} \\
  \cmidrule{5-8}          & \multicolumn{1}{c}{} &       &       &  \multicolumn{1}{p{8em}}{Sequence} & Data$\dagger$ & Training$\dagger$ & Testing$\dagger$ &       & \multicolumn{1}{c}{} & \multicolumn{1}{c}{} &  \\
      \midrule
      SCD   & Sunnybrook Cardiac Data (Cardiac MR Left Ventricle Segmentation Challenge data ) & 2009  & \multicolumn{1}{p{11.875em}}{\cite{Radau_Lu_Connelly_Paul_Dick_Wright2009}} & SSFP & 45    & 45    & -     & LV, MYO & \multicolumn{1}{l}{HF, HCM, NHS} & \multicolumn{1}{l}{GE Signa 1.5T} & MICCAI 2009 \\
      LVSC  & LV Segmentation Challenge & 2011  & \cite{LVSC} & SAX and LAX SSFP & 200   & 100   & 100   & LV, MYO & \multicolumn{1}{l}{MI} & GE Signa 1.5T, Philips Achieva (1.5T, 3.0T, and Intera 1.5T), and Siemens\newline{}(Avanto 1.5T, Espree 1.5T and Symphony 1.5T) & MICCAI 2011 \\
      RVSC  & Right Ventricle Segmentation Challenge & 2012  & \cite{petitjean_right_2015} & bSSFP & 48    & 16    & 32    & \multicolumn{1}{p{12.54em}}{RV} & N/A   & Siemens Symphony\newline{}Tim 1.5T & MICCAI 2012 \\
      LASC'13:  & Left Atrial Segmentation Challenge & 2013  & \cite{7029623} & bSSFP & 30  & 10    & 20    & \multicolumn{1}{p{12.54em}}{LA, PV} & N/A   & Philips Achieva 1.5T & MICCAI 2013 \\
      MM-WHS & Multi-Modality Whole Heart Segmentation & 2017  & \cite{MMWHS} & bSSFP & 60 & 20    & 40    & LV, RV, LA, RA, MYO, AO & \multicolumn{1}{l}{DCM} & Philips 1.5T, Siemens Magnetom Avanto 1.5T & MICCAI 2017 \\
      ACDC  & Automated Cardiac Diagnosis Challenge  & 2017  & \cite{ACDC} & $1.5 T$ and $3 T$, SSFP & 150   & 100   & 50    & \multicolumn{1}{p{12.54em}}{LV, RV, MYO} & NOR, MI, HCM, ARV & Siemens Area and Trio Tim & MICCAI 2017 \\
      ASC   & Atria Segmentation Challenge & 2018  & \cite{ASC} & LGE & 154   & 100   & 54    & \multicolumn{1}{p{12.54em}}{LA} & AF    & \multicolumn{1}{l}{GE-MRIs} & MICCAI 2018 \\
      MSCMR & Multi-sequence Cardiac MR Segmentation Challenge & 2019  & \cite{8458220} & LGE, T2 and bSSFP & 115   & 75    & 40    & LV, RV, MYO & \multicolumn{1}{l}{MI, DCM} & \multicolumn{1}{l}{Philips Achieva 1.5 T} & MICCAI 2019 \\
      M\&Ms-1 & Multi-Centre, Multi-Vendor \& Multi-Disease Cardiac Image Segmentation Challenge & 2020  & \cite{9458279} & N/A   & 375   & 175   & 200   & LV, RV, MYO & \multicolumn{1}{l}{MH, DCM, NHS} & Siemens MAGNETOM \newline{}(Avanto and Skyra)\newline{}Philips Achieva\newline{}GE Signa Excite\newline{}Canon Vantage Orian\newline{} & MICCAI 2020 \\
      MyoPS 20 & Myocardial pathology segmentation combining multi-sequence CMR & 2020  & \cite{MyoPS} & bSSFP, LGE, T2 CMR & 45 & 25  & 20 & \multicolumn{1}{p{12.54em}}{MYO Scar, MYO Edema} & MI    & Philips Achieva 1.5T & MICCAI 2020 \\
      M\&Ms-2 & Multi-Disease, Multi-View \& Multi-Center Right Ventricular Segmentation in Cardiac MRI & 2020  & \cite{10103611} & N/A   & 360   & 200   & 160   & LV, RV, MYO & DCM, HCM, CAM, ARR, TF,  IC, TR, NHS & Siemens, General Electric, and Philips & MICCAI 2021 \\
      \bottomrule
      \end{tabular}%
    }
  \label{tab:related_challenges}%
\end{table*}%

To curate an extreme CMR dataset with respiratory motion artifacts, one way is to screen retrospective images stored in the hospital imaging database and identify those problematic ones. This requires considerable human efforts for screening and it is often restricted by data governance regulations. It may also introduce confounding factors such as vendors, scan protocols, and pathologies that are difficult to disentangle from respiratory motion. Instead, we design a prospective study in which healthy volunteers are recruited to perform different breath-hold behaviors during one imaging visit using the same scanner. As the confounding factors of MRI equipment and scan protocols are controlled, the curated CMR dataset is established in specific to respiratory motion artifacts. This manuscript is structured following the Biomedical Image Analysis Challenges (BIAS) guideline~\citep{MAIERHEIN2020101796}, and it provides a comprehensive summary of the \textit{CMRxMotion Challenge}, including the data acquisition protocol, data annotation procedures, evaluation tasks and metrics, ranking methodology, award scheme, and challenge results.

\section{Related work}
Over the past two decades, a multitude of challenges focused on cardiac chamber segmentation have been instrumental in advancing the field. These events provide a standardized platform for the objective comparison of algorithms, fostering innovation in clinical and research applications. This section reviews the evolution of these challenges to contextualize the unique contribution of \textit{CMRxMotion}, which, as summarized in \Cref{tab:related_challenges}, is the first to specifically address the robustness of segmentation models to respiratory motion artifacts.

Early challenges primarily targeted the segmentation of core cardiac structures in both healthy and pathological cases, namely the left ventricle (LV), right ventricle (RV), and left ventricular myocardium (MYO). For instance, given the clinical importance of the LV ejection fraction (EF) as a predictor of heart disease~\citep{LVEF}, foundational challenges such as the Sunnybrook Cardiac Data (SCD)~\citep{Radau_Lu_Connelly_Paul_Dick_Wright2009} and the LV Segmentation Challenge (LVSC)~\citep{LVSC} centered on the LV and MYO. Similarly, to address the need for automated evaluation of the RV in conditions like dysplasia and cardiomyopathy~\citep{10.1001/jamacardio.2016.5034}, the Right Ventricle Segmentation Challenge (RVSC) was established to benchmark methods for delineating the RV endocardium and epicardium~\citep{petitjean_right_2015}.

As the field matured, the scope of these challenges expanded significantly. The focus shifted towards more complex and clinically relevant problems, including multi-structure segmentation, cross-domain generalization, and the analysis of pathological tissue. Challenges began to incorporate a wider range of anatomical targets, such as the atria~\citep{7029623, ASC} and, in some cases, the entire heart~\citep{MMWHS}. Concurrently, a major focus became assessing model robustness to data heterogeneity. This was addressed by curating datasets from multiple institutions, featuring different scanner vendors, imaging protocols, and diverse patient populations~\citep{8458220, 9458279, 10103611}. More recently, challenges have also begun to target the direct segmentation of pathological tissue, such as myocardial scar and edema in the MyoPS challenge~\citep{MyoPS}.

Despite this progress in tackling data heterogeneity, the challenge landscape has largely overlooked the critical problem of image quality degradation from patient motion, a factor that frequently compromises the clinical utility of automated methods. In clinical practice, poor breath-holding is a common occurrence, leading to artifacts that degrade image quality, blur anatomical boundaries, and introduce heterogeneous intensity distributions~\citep{wang2021joint}. Such artifacts can severely compromise the performance of segmentation models trained exclusively on high-quality data, leading to inaccurate analysis and potentially erroneous clinical conclusions. To date, no challenge has provided a framework to \textit{stress test} algorithms against this common source of failure. Therefore, evaluating model performance in the presence of motion artifacts is essential for developing truly robust and clinically trustworthy tools~\citep{paschali2018generalizability}.

To address this critical gap, we organized the \textit{CMRxMotion} challenge. Its primary goal was to establish the first benchmark for a twofold problem: first, the automated assessment of CMR image quality itself, and second, the evaluation of segmentation model robustness in the presence of motion artifacts. By providing a curated dataset with controlled motion degradation, this challenge aims to drive the development of reliable analysis tools that are resilient to the image quality issues frequently encountered in real-world clinical practice.

\section{Materials and methods}

\subsection{Data acquisition}
The study received ethical approval from the institutional review board of Fudan University (FE20017), and all participants provided written informed consent. The cohort consisted of 40 healthy volunteers (24 male, 16 female) with an age range of 19 to 48 years and a bodyweight range of 45 to 85 kg.
All imaging was performed on a 3T MRI scanner (Siemens MAGNETOM Vida) at the Zhangjiang International Brain Imaging Center, Fudan University. We employed a clinical TrueFISP balanced Steady-State Free Precession (bSSFP) cine sequence, following the protocols detailed in our previous work~\citep{wang2021recommendation}. For this challenge, we released short-axis (SAX) image volumes at the end-diastolic (ED) and end-systolic (ES) frames. The acquisition parameters were set to an in-plane spatial resolution of $2.0 \times 2.0$ mm, a slice thickness of $8.0$ mm, and a slice gap of $4.0$ mm.

\begin{figure*}[hbtp]
  \centering
  \includegraphics[width=\linewidth]{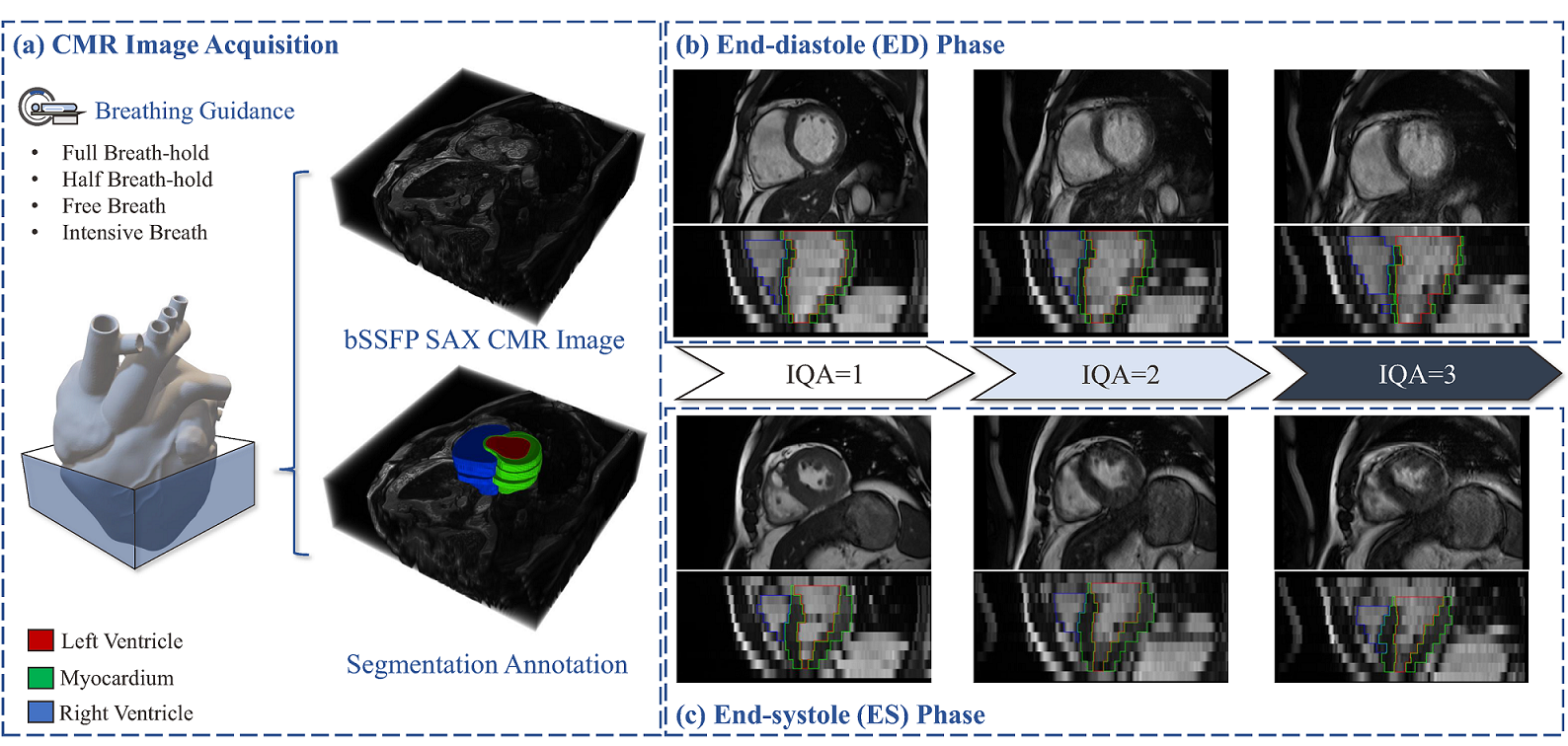}
  \caption{Visualization example of paired short-axis (SAX) CMR images with different Image Quality Assessment (IQA) scores, acquired from the same volunteer performing four distinct breath-hold behaviors following specific breathing guidance during CMR acquisition. (b) End-diastolic (ED) images, showing three images with progressively increasing IQA scores, are presented in the top row. (c) End-systolic (ES) images, which also display three CMR scans in both axial and coronal planes, are presented in the bottom row.}
  \label{fig1}
\end{figure*}

To generate a dataset with a controlled spectrum of image quality, we intentionally deviated from the standard operating procedure (SOP) for breath-holding. Each volunteer underwent four separate scans during a single imaging session, each with a different breathing instruction: a) adhere to the breath-hold requirements; b) halve the breath-hold period; c) breathe freely; and d) breathe intensively.  This protocol yielded a grouped set of CMR images for each volunteer, capturing a range of motion artifact severity from none (high-quality) to severe. From each of the four acquisitions, 3D image volumes at the ED and ES frames were extracted, resulting in a total of eight volumes per volunteer and 320 volumes for the entire dataset. \Cref{fig1} demonstrates an example of the resulting motion-degraded images.

\subsection{Preprocessing}
All acquired short-axis cine images were processed through a standardized pipeline. First, the DICOM files were anonymized and subsequently converted to the NIfTI format using the \textit{dicom2nifti} Python package\footnote{\href{https://dicom2nifti.readthedocs.io}{https://dicom2nifti.readthedocs.io/en/latest/}}. From each 4D cine sequence, the 3D image volumes corresponding to the end-diastolic (ED) and end-systolic (ES) frames were extracted. Finally, to ensure compatibility with various NIfTI viewers such as ITK-SNAP, orientation information was removed from the image headers. This process yielded a total of 320 3D image volumes for the challenge dataset.

\subsection{Challenge dataset}
The 320 image volumes were partitioned at the subject level to prevent data leakage between sets. The dataset was randomly split as follows: a \textbf{training set} of 160 volumes (20 volunteers), a \textbf{validation set} of 40 volumes (5 volunteers), and a \textbf{test set} of 120 volumes (15 volunteers). The training data, along with corresponding ground truth annotations, were released to all registered participants. The validation images were also released, but their annotations were withheld to power a live leaderboard for online evaluation. Both the images and annotations for the test set were fully withheld and used exclusively for the final, offline evaluation of submitted containerized algorithms. All annotations were manually delineated by radiologists with over six years of experience. Following the conclusion of the challenge, the data and annotations remain publicly accessible under a CC-BY non-commercial license to encourage further research.

\begin{table}[htbp]
  \centering
  \caption{Data split and percentage of Image Quality Assessment (IQA) label among different data split. Data in the IQA label proportion column represents subjects(percentage).}
    \resizebox{\linewidth}{!}{
    \begin{tabular}{cccccc}
    \toprule
    \multicolumn{1}{c}{\multirow{2}[4]{*}{\textbf{Data}}} & \multicolumn{1}{c}{\multirow{2}[4]{*}{\textbf{Subjects}}} & \multicolumn{1}{c}{\multirow{2}[4]{*}{\textbf{Image cases}}} & \multicolumn{3}{c}{\textbf{IQA Label Proportion}} \\
\cmidrule{4-6}          &       &       & \textbf{1} & \textbf{2} & \textbf{3} \\
    \midrule
    Training Set & 20    & 160   & 48 (32\%) & 54 (36\%) & 54 (32\%) \\
    Validation Set & 5     & 40    & 18 (45\%) & 18 (45\%) & 4 (10\%) \\
    Test Set & 15    & 120   & 48 (40\%) & 54 (45\%) & 18 (15\%) \\
    \bottomrule
    \end{tabular}%
  }
  \label{tab:IQA_distribution}%
\end{table}%

\begin{table*}[!hbtp]
  \centering
  \caption{Assessment criteria for subjective image quality and label definition in Task 1.}
  \resizebox{\linewidth}{!}
  {
    \begin{tabular}{>{\centering}p{9em}p{15em}p{20em}c}
    \toprule
    \textbf{5-point Likert Scale} & \textbf{Subjective Image Quality Grade} & \textbf{Details} & \multicolumn{1}{>{\centering}p{15.125em}}{\textbf{IQA Label}} \\
    \midrule
    V     & Excellent & No artifacts present & \multirow{2}[2]{*}{\textbf{1}} \\
    IV    & More than adequate for diagnosis & Minor artifacts present but image quality somewhat reduced &  \\
    \midrule
    III   & Adequate for diagnosis & Minor artifacts present and image quality somewhat reduced but still sufficient for diagnosis & \textbf{2} \\
    \midrule
    II    & Questionable for diagnosis & Image quality impaired by artifacts so diagnostic value of images is questionable & \multirow{2}[2]{*}{\textbf{3}} \\
    I     & Non-diagnostic & Image quality heavily impaired by artifacts and readers not able to assess &  \\
    \bottomrule
    \end{tabular}%
  }
  \label{tab:IQA_Label}%
\end{table*}

\section{Challenge design}
The \textit{CMRxMotion challenge} was organized by a collaborative team of researchers from Fudan University, the University of Edinburgh, Imperial College London, the University of Oxford, and Zhongshan Hospital. The challenge was designed to establish a benchmark for two critical, unsolved problems in automated CMR analysis: the image quality assessment (IQA) and the robust cardiac segmentation (RCS)  in the presence of respiratory motion. 

\subsection{Task 1: automated image quality assessment}
\subsubsection{Motivation}
CMR images that are severely degraded by motion artifacts are often diagnostically unacceptable and may necessitate re-acquisition, leading to increased scan time and patient discomfort. The development of an automated IQA model capable of identifying such low-quality images could provide immediate, real-time feedback to MR technologists, improving clinical workflow efficiency. The primary objective of Task 1 was, therefore, to benchmark algorithms for the automated evaluation of respiratory motion artifacts in CMR images.

\subsubsection{Annotation protocol}
\label{sec:annotation_t1}
The image quality of all 320 volumes was manually assessed by two experienced radiologists (Z.S. and C.D.) using 3D Slicer\footnote{\href{http://www.slicer.org}{http://www.slicer.org}}. Following a consensus meeting to establish a standardized protocol.  The standard 5-point Likert scale is defined as follows: excellent diagnostic quality (5), more than adequate for diagnosis (4), adequate for diagnosis (3), questionable for diagnosis (2), and non-diagnostic (1). For better reproducibility, three levels of motion artifacts are defined based on the original 5-point scores. For the purpose of creating a well-defined classification task, these five scores were consolidated into three final labels, as detailed in \Cref{tab:IQA_Label}. Images with scores of 4 or 5 were categorized as having \textbf{mild motion} (Label 1), images with a score of 3 were categorized as having \textbf{intermediate motion} (Label 2), and images with scores of 1 or 2 were categorized as having \textbf{severe motion} (Label 3). The objective for participants was to develop a model that predicts one of these three labels for a given input image volume.

\subsubsection{Evaluation metrics}
Cohen's Kappa statistics is used to evaluate the model performance in this task. It is a common metric to measure the level of agreement between two raters who classify items into mutually exclusive categories~\citep{benchoufi2020interobserver}:

\begin{equation}
    \kappa = \frac{p_o-p_e}{1-p_e}
    \label{kappa}
\end{equation}
    
where $p_o$ is the observed agreement ratio (overall accuracy of the model) and $p_e$ is the agreement between the prediction and the ground truth as if happening by chance. Participating teams submitted a containerized Docker image designed to predict a label from the set \{1, 2, 3\}, corresponding to mild, intermediate, and severe motion artifacts. We calculate the weighted Cohen's Kappa between the submission and the manual annotation. 

For a more detailed performance analysis, we also computed several secondary metrics: Accuracy, Precision, and Recall. Accuracy measures the overall proportion of correctly classified samples:
\begin{equation}
  \text{Accuracy} = \frac{1}{N} \sum_{i=1}^{N} \mathbf{1} \left\{ \hat{y}_i = y_i \right\}
  \label{eq:acc}
\end{equation}
where $N$ is the total number of samples, and $\hat{y}_{i}$ and $y_i$ are the predicted and true labels. $\mathbf{1}\{\cdot\}$ represents the indicator function, which equals 1 if the condition is true, and 0 otherwise.

Precision and Recall were calculated on a per-class basis. Precision for class $k$ is the ratio of true positive predictions to the total number of instances predicted as class $k$:
\begin{equation}
  \text{Precision}_k = \frac{TP_k}{TP_k + FP_k}
\end{equation}
where $TP_k$ is the number of true positives and $FP_k$ is the number of false positives for class $k$. Recall for class $k$ is the proportion of true positives among all samples that truly belong to class $k$:
\begin{equation}
  \text{Recall}_k = \frac{TP_k}{TP_k + FN_k}
\end{equation}
where $FN_k$ is the number of false negatives for class $k$. It is important to note that these secondary metrics were used for descriptive analysis only and did not contribute to the final ranking. All metrics were calculated using the \textit{scikit-learn} Python package\footnote{\href{https://scikit-learn.org/stable/modules/generated/sklearn.metrics.cohen_kappa_score.html}{https://scikit-learn.org/}}.

\subsubsection{Ranking scheme}
The final ranking on the test set was generated by sorting the submissions based on their Cohen's Kappa scores. To evaluate whether the performance differences between algorithms were statistically significant, we performed bootstrap sampling to calculate the 95\% confidence intervals (95\% CI) for each submission's score.

\subsection{Task 2: robust cardiac segmentation}
\subsubsection{Motivation} 
Learning-based segmentation models are known to be susceptible to failure when encountering images with quality characteristics unseen during training. The objective of Task 2 was to challenge participants to develop segmentation models that are robust to the presence of respiratory motion artifacts.

\subsubsection{Annotation protocol}
\label{sec:annotation_t2}
The segmentation annotation protocol followed the standards of previous CMR challenges. Manual contours were delineated for the LV and RV blood pools and the MYO. Critically, segmentation masks were only generated for images deemed to be of diagnostic quality (i.e., those with an IQA score of 1 or 2 from Task 1). The annotations were performed by an experienced technician (X.K.) using 3D Slicer and subsequently reviewed and refined by two senior radiologists (Z.S. and C.D.). The final segmentation masks used the following label definitions: 0 (Background), 1 (LV), 2 (MYO), and 3 (RV). \Cref{fig2} provides a visualization of the segmentation labels.

\subsubsection{Evaluation metrics}
Segmentation performance was evaluated using two standard metrics: the Dice Similarity Coefficient (DSC) and the 95th percentile of the Hausdorff Distance (HD95). The DSC measures the volumetric overlap between the ground truth ($X$) and the predicted segmentation ($Y$):
\begin{equation}
    \text{DSC} = \frac{2|X \cap Y|}{|X|+|Y|}
    \label{Dice}
\end{equation}
The HD95 measures the 95th percentile of the maximum surface distance between the contours of the two masks, providing a robust measure of boundary accuracy:
\begin{equation}
        \text{HD95} = max\{\mathop{P_{95\%}}_{x\in X}d(x, Y), \mathop{P_{95\%}}_{y\in Y}d(y, X)\}
        \label{HD95}
\end{equation}
where $d(x,Y)=\min_{y\in Y} {\|}x-y{\|}$. Both metrics were calculated using the MedPy library\footnote{\href{http://loli.github.io/medpy/}{http://loli.github.io/medpy/}}.

\begin{center}
  \centering
  \includegraphics[width=\linewidth]{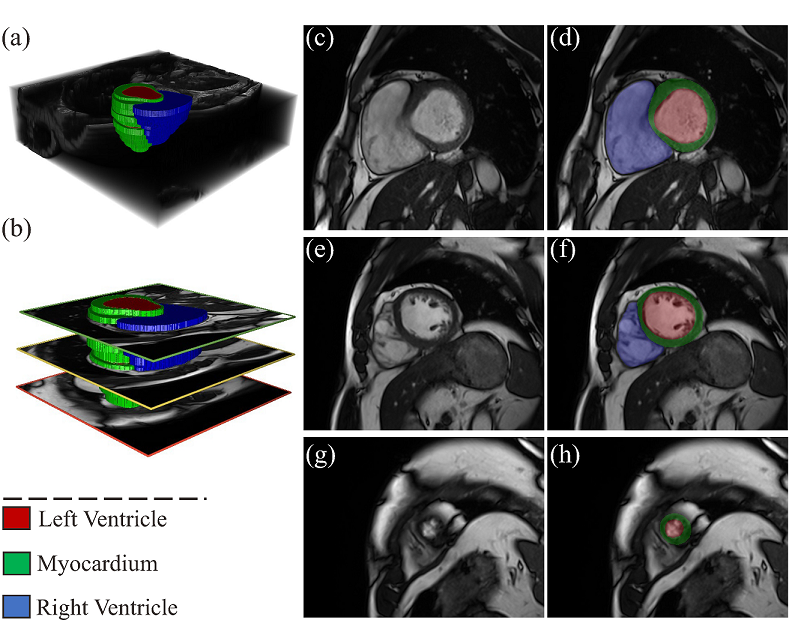}
  \captionof{figure}{Visualization of segmentation labels for Task 2. (a) Volume rendering of a short-axis CMR image. (b) Volume rendering of the corresponding segmentation mask. (c-h) Paired image slices and segmentation masks for the basal, mid-ventricular, and apical regions. The LV is shown in red, MYO in green, and RV in blue.}
  \label{fig2}
\end{center}

\subsubsection{Ranking scheme}
To determine the final ranking, we employed a \textit{rank-then-aggregate} strategy. For each case in the test set, we first ranked all participating teams based on their DSC and HD95 scores for each of the three structures (LV, MYO, and RV). This process resulted in six individual rankings per case for each team. The final ranking score for each team was then calculated as the average of all these individual ranks across all test cases. This method ensures that the final ranking reflects consistent performance across all structures and metrics, rather than being skewed by exceptional performance on a single metric. The Wilcoxon signed-rank test was used to assess the statistical significance of differences between the final ranking scores of the teams.

\subsection{Rules and timeline schedule}
\subsubsection{Rules}
\label{sec:rule}
The challenge was governed by a set of rules designed to ensure a fair, transparent, and reproducible evaluation.
\begin{enumerate}
     \item \textbf{Submission policy}: The challenge consisted of two main phases. During the validation phase, participants were permitted to make up to three submissions per day to a live leaderboard for each task. For the final testing phase, each team was allowed a single submission of a containerized Docker image for the official offline evaluation. Besides, members of the organizers could not participate in the challenge.
    \item \textbf{Use of external data}: Teams were permitted to use publicly available external data for model training. However, all external data sources were required to be explicitly cited and described in the accompanying methodology paper. This rule was intended to encourage the exploration of techniques such as transfer learning and domain adaptation.
    \item \textbf{Reproducibility}: To ensure the replicability of results, participants were required to submit a manuscript detailing their methods. For the final testing phase, a self-contained Docker image, including all source code and pretrained model weights, was mandatory. Public release of the code was strongly encouraged to benefit the wider research community.
    \item \textbf{Automation and fairness}: All submitted algorithms had to be fully automatic, with no manual intervention permitted during inference. To ensure a standardized and fair evaluation environment, network access was strictly prohibited during the execution of the Docker containers on the test set.
\end{enumerate}

\subsubsection{Timeline}
The \textit{CMRxMotion} challenge was a one-time event, with foundational work (including data collection, processing, annotation, and visual inspection) conducted between May 2021 and April 2022. During this period, the challenge proposal was formulated and submitted, receiving official acceptance from the \textit{MICCAI 2022} Satellite Events Committee on February 28, 2022. The public phase of the competition began with the launch of the official website on April 1, 2022, with participant registration opening shortly after on April 15. The training dataset—complete with images and ground truth labels—along with a brief challenge description paper~\citep{2022arXiv221006385W}, was released to all registered teams on May 31. The validation phase commenced on July 1 with the release of the validation images and the opening of the live leaderboard on the Synapse platform\footnote{\href{https://www.synapse.org/\#!Synapse:syn32407769/wiki/}{https://www.synapse.org/\#!Synapse:syn32407769/wiki/}}. In parallel, the test set was annotated by three senior radiologists. The final testing phase was initiated on September 1 with the launch of the Docker submission platform for offline evaluation. The challenge culminated at the \textit{MICCAI 2022} conference. Top-performing teams were invited to submit full papers, which, after peer review, were published in the \textit{STACOM} workshop proceedings. The top three winning teams for each task presented their work at the workshop held on September 18, 2022, where they were officially recognized and awarded.

\subsubsection{Infrastructure}
The challenge was supported by a comprehensive technical infrastructure distributed across several platforms, each serving a specific function:
\begin{enumerate}
  \item \textbf{Challenge homepage}: The official website (\url{http://cmr.miccai.cloud}) served as the central information hub for participant registration, rules, timelines, and data usage agreements.
    
  \item \textbf{Validation platform}: The Synapse platform (\url{https://www.synapse.org/#!Synapse:syn32407769/wiki/}) was used for secure data distribution and to host a live leaderboard for real-time, automated evaluation of submissions.

  \item \textbf{Testing platform}: A dedicated platform (\url{https://docker.miccai.cloud}) managed the collection and execution of containerized Docker submissions on a standardized GPU environment for the final, offline evaluation.

  \item \textbf{Public archives}: To ensure long-term accessibility, all resources are publicly archived. A GitHub repository (\url{https://github.com/CMRxMotion}) provides access to evaluation scripts and code, while the submitted Docker images are archived on Docker Hub (\url{https://hub.docker.com/repositories/cmrxmotion}) under an Apache 2.0 license.
\end{enumerate}

\section{Submission and evaluation}
The evaluation procedure for the \textit{CMRxMotion challenge} was conducted in two distinct stages: a validation phase with a live leaderboard, and a final testing phase based on containerized algorithm submissions. This two-stage design allowed for iterative development while ensuring a final, definitive ranking on unseen data.

\subsection{Validation phase}
During the validation phase, participants could benchmark their models using the provided training set (160 volumes from 20 volunteers) and submit predictions for the validation set (40 volumes from 5 volunteers). To facilitate iterative refinement, teams were permitted up to three submissions per day to a live leaderboard for each task.

For the \textbf{IQA task}, submissions were in the form of a comma-separated values (CSV) file containing the predicted quality label \{1, 2, 3\} for each validation case. An automated backend calculated the Cohen's Kappa score against the withheld ground truth and updated the leaderboard. For the \textbf{RCS task}, participants submitted a compressed archive of their predicted segmentation masks in NIfTI format. The backend computed the Dice Similarity Coefficient (DSC) and 95\% Hausdorff Distance (HD95) for the LV, MYO, and RV. It is important to note that images with severe motion artifacts (IQA Label 3) were excluded from the segmentation evaluation due to the lack of reliable ground truth annotations.

\begin{table*}[htbp]
  \centering
  \caption{Overview of all workshop papers accepted in the STACOM proceedings, along with the publicly available or otherwise accessible reproducible resources provided by the corresponding participating teams. TID denotes the team identifier for Task 1 (TID prefixed with A) and Task 2 (TID prefixed with S). Checkmark (\checkmark) indicates that the team participated in the corresponding task or has made the relevant resources available.
  }
  \resizebox{\linewidth}{!}{
    \begin{tabular}{lllccccccccp{40.665em}}
    \toprule
    \multicolumn{1}{l}{\multirow{2}[3]{*}{\textbf{TID}}} & \multicolumn{1}{l}{\multirow{2}[3]{*}{\textbf{Team}}} & \multicolumn{1}{l}{\multirow{2}[3]{*}{\textbf{References}}} & \multicolumn{2}{c}{\textbf{Participating}} &       & \multicolumn{2}{c}{\textbf{Source Code}} &       & \multicolumn{2}{c}{\textbf{Docker Image}} & \multirow{2}[3]{*}{\textbf{Publicly Accessible and Reproducible Resources}} \\
\cmidrule{4-5}\cmidrule{7-8}\cmidrule{10-11}          &       &       & \multicolumn{1}{c}{\textbf{Task1}} & \multicolumn{1}{c}{\textbf{Task2}} &       & \multicolumn{1}{c}{\textbf{Task1}} & \multicolumn{1}{c}{\textbf{Task2}} &       & \multicolumn{1}{c}{\textbf{Task1}} & \multicolumn{1}{c}{\textbf{Task2}} & \multicolumn{1}{c}{} \\
    \midrule
    \textbf{A1}    & UON\_IMA & \cite{li-motion-related-2022} & \checkmark     &     &       & \checkmark     &     &       & \checkmark     &     & • \textbf{Source Code}:  \textcolor[rgb]{ .02,  .388,  .757}{\url{https://github.com/ruizhe-l/CMRxMotion}} \newline{}• Docker Image: \textcolor[rgb]{ .02,  .388,  .757}{\url{https://hub.docker.com/r/cmrxmotion/a1.uon_ima}} \\
    \textbf{S2}    & Med-Air & \cite{gong-robust-2022} &     & \checkmark     &       &     & \checkmark     &       &     & \checkmark     & • \textbf{Source Code}: \textcolor[rgb]{ .02,  .388,  .757}{https://github.com/CMRxMotion/S2.Med-Air} \newline{}• Docker Image: \textcolor[rgb]{ .02,  .388,  .757}{\url{https://hub.docker.com/r/cmrxmotion/s2.med-air}} \\
    \textbf{A2/S7} & Philips\_CTS & \cite{li-automatic-2022}  & \checkmark     & \checkmark     &       & \checkmark     & \checkmark     &       & \checkmark     & \checkmark     & • \textbf{Source Code}: \textcolor[rgb]{ .02,  .388,  .757}{\url{https://github.com/CMRxMotion/A2_S7.Philips_CTS}} \newline{}• Docker Image 1: \textcolor[rgb]{ .02,  .388,  .757}{\url{https://hub.docker.com/r/cmrxmotion/a2.philips_cts}} \newline{}• Docker Image 2: \textcolor[rgb]{ .02,  .388,  .757}{\url{https://hub.docker.com/r/cmrxmotion/s7.philips_cts}} \\
    \textbf{A3/S5} & OpenGTN & \cite{amirrajab-cardiac-2022} & \checkmark     & \checkmark     &       & \checkmark     & \checkmark     &       & \checkmark     & \checkmark     & • \textbf{Source Code}: \textcolor[rgb]{ .02,  .388,  .757}{\url{https://github.com/CMRxMotion/A3_S5.OpenGTN}} \newline{}• Docker Image 1:  \textcolor[rgb]{ .02,  .388,  .757}{\url{https://hub.docker.com/r/cmrxmotion/a3.opengtn}}  \newline{}• Docker Image 2:  \textcolor[rgb]{ .02,  .388,  .757}{\url{https://hub.docker.com/r/cmrxmotion/s5.opengtn}}\\
    \textbf{A4}    & issun & \cite{sun_combination_2022} & \checkmark     &     &       & \checkmark     &     &       & \checkmark     &     & • \textbf{Source Code}: \textcolor[rgb]{ .02,  .388,  .757}{\url{https://github.com/CMRxMotion/A4.issun}} \newline{}• Docker Image: \textcolor[rgb]{ .02,  .388,  .757}{\url{https://hub.docker.com/r/cmrxmotion/a4.issun}} \\
    \textbf{A5/S4} & Tewodrosw & \cite{arega-automatic-2022} & \checkmark     & \checkmark     &       & \checkmark     & \checkmark     &       & \checkmark     & \checkmark     & • \textbf{Source Code}: \textcolor[rgb]{ .02,  .388,  .757}{\url{https://github.com/tewodrosweldebirhan/cardiac_segmentation_cmrxmotion2022}} \newline{}• Docker Image: \textcolor[rgb]{ .02,  .388,  .757}{\url{https://hub.docker.com/r/cmrxmotion/s4.tewodrosw}} \\
    \textbf{A6/S3} & CMR.Love.LHND & \cite{yang-deep-2022} & \checkmark     & \checkmark     &       & \checkmark     & \checkmark     &       & \checkmark     & \checkmark     & • \textbf{Source Code}:\textcolor[rgb]{ .02,  .388,  .757}{\url{https://github.com/CMRxMotion/A6_S3.CMR.Love.LHND}} \newline{}• Docker Image 1: \textcolor[rgb]{ .02,  .388,  .757}{\url{https://hub.docker.com/r/cmrxmotion/a6.cmr.love.lhnd}} \newline{}• Docker Image 2: \textcolor[rgb]{ .02,  .388,  .757}{\url{https://hub.docker.com/r/cmrxmotion/s3.cmr.love.lhnd}}\\
    \textbf{A7/S8} & MI-IST\_DA & \cite{ranem_detecting_2022} & \checkmark     & \checkmark     &       & \checkmark     & \checkmark     &       & \checkmark     & \checkmark     & • \textbf{Source Code}: \textcolor[rgb]{ .02,  .388,  .757}{\url{https://github.com/MECLabTUDA/QA_med_data/tree/QMRxMotion}} \newline{}• Docker Image 1:  \textcolor[rgb]{ .02,  .388,  .757}{\url{https://hub.docker.com/r/cmrxmotion/a7.mi-ist_da}} \newline{}• Docker Image 2:  \textcolor[rgb]{ .02,  .388,  .757}{\url{https://hub.docker.com/r/cmrxmotion/s8.mi-ist_da}}\\
    \textbf{A8/S1} & UA-SVCC & \cite{mora-rubio-deep-2022} & \checkmark     & \checkmark     &       & \checkmark     & \checkmark     &       & \checkmark     & \checkmark     & • \textbf{Source Code}: \textcolor[rgb]{ .02,  .388,  .757}{\url{https://github.com/CMRxMotion/A8_S1.UA-SVCC}} \newline{}• Docker Image 1: \textcolor[rgb]{ .02,  .388,  .757}{\url{https://hub.docker.com/r/cmrxmotion/a8.ua-svcc}} \newline{}• Docker Image 2: \textcolor[rgb]{ .02,  .388,  .757}{\url{https://hub.docker.com/r/cmrxmotion/s1.ua-svcc}} \\
    \textbf{A9/S6} & Abdul & \cite{qayyum_automatic_2022} & \checkmark     & \checkmark     &       & \checkmark     & \checkmark     &       & \checkmark     & \checkmark     & • \textbf{Source Code}: \textcolor[rgb]{ .02,  .388,  .757}{\url{https://github.com/RespectKnowledge/CMRxMotion_Solution}} \newline{}• Docker Image 1: \textcolor[rgb]{ .02,  .388,  .757}{\url{https://hub.docker.com/r/cmrxmotion/a9.abdul}} \newline{}• Docker Image 2: \textcolor[rgb]{ .02,  .388,  .757}{\url{https://hub.docker.com/r/cmrxmotion/s6.abdul}} \\
    \textbf{S9}   & ML-Labs & \cite{garcia-cabrera_cardiac_2022} &     & \checkmark     &       &     & \checkmark     &       &     & \checkmark     & • \textbf{Source Code}: \textcolor[rgb]{ .02,  .388,  .757}{\url{https://github.com/CMRxMotion/S9.ML-Labs}} \newline{}• Docker Image: \textcolor[rgb]{ .02,  .388,  .757}{\url{https://hub.docker.com/r/cmrxmotion/s9.ml-labs}} \\
    \textbf{A10/S10} & Sano  & \cite{grzeszczyk_multi-task_2022} & \checkmark     & \checkmark     &       & \checkmark     & \checkmark     &       & \checkmark     & \checkmark     & • \textbf{Source Code}: \textcolor[rgb]{ .02,  .388,  .757}{\url{https://github.com/CMRxMotion/A10_S10.Sano}} \newline{}• Docker Image 1: \textcolor[rgb]{ .02,  .388,  .757}{\url{https://hub.docker.com/r/cmrxmotion/a10.sano}} \newline{}• Docker Image 2: \textcolor[rgb]{ .02,  .388,  .757}{\url{https://hub.docker.com/r/cmrxmotion/s10.sano}} \\
    \textbf{S11}    & HAHA2 & \cite{ma_semi-supervised_2022} &     & \checkmark     &       &     & \checkmark     &       &     & \checkmark     & • \textbf{Source Code}: \textcolor[rgb]{ .02,  .388,  .757}{\url{https://github.com/MAWanqin2002/STACOM2022Ma}} \newline{}• Docker Image: \textcolor[rgb]{ .02,  .388,  .757}{\url{https://hub.docker.com/r/cmrxmotion/s11.haha2}} \\
    \textbf{S12}   & sots  & \cite{kou_3d_2022} &     & \checkmark     &       &     & \checkmark     &       &     & \checkmark     & • \textbf{Source Code}: \textcolor[rgb]{ .02,  .388,  .757}{\url{https://github.com/CMRxMotion/S12.sots}}  \newline{}• Docker Image: \textcolor[rgb]{ .02,  .388,  .757}{\url{https://hub.docker.com/r/cmrxmotion/s12.sots}} \\
    \bottomrule
    \end{tabular}%
  }
  \label{tab:overview}%
\end{table*}%

\subsection{Testing phase}
The final ranking was determined through an offline evaluation on a sequestered test set (120 volumes from 15 volunteers). For this phase, each team submitted a self-contained Docker image of their algorithm, including all source code and pretrained model weights. This approach ensured reproducibility and prevented overfitting to the validation set leaderboard. All submissions were executed in a standardized, isolated environment on a GPU server with the following specifications: 10 CPU cores (2.50 GHz), 32 GB RAM, and an NVIDIA Tesla V100 GPU (32 GB VRAM). To ensure fairness, network access was disabled, and a maximum inference time of four hours was allotted for each task.

The final ranking for the \textbf{IQA task} was determined by sorting submissions in descending order of their Cohen's Kappa score on the test set. For the \textbf{RCS task}, the final ranking was based on a robust \textit{rank-then-aggregate} strategy. For each test case, we ranked all teams based on their DSC and HD95 scores for each of the three structures (LV, MYO, and RV). The final score for each team was the average of these individual ranks across all cases. This method, which aligns with conventions established in previous segmentation challenges (e.g.,~\citep{maier_isles_2017, baid2021rsnaasnrmiccai}), rewards consistently high performance across all metrics and anatomical structures. The Wilcoxon signed-rank test was subsequently employed to assess whether the differences in final ranking scores between teams were statistically significant—that is, whether the higher-ranked methods demonstrated a statistically significant advantage over the lower-ranked ones.

\section{Results}
The \textit{CMRxMotion challenge} garnered substantial interest from the research community, with formally 112 registered participants and a total of 275 and 340 submissions to the validation leaderboard for Task 1 and Task 2, respectively. For the final testing phase, 10 teams submitted Docker images for the IQA task, and 13 teams submitted for the RCS task. The top-3 teams were invited to present their work at the \textit{STACOM} satellite workshop at \textit{MICCAI 2022}, and 14 methodology papers were accepted into the workshop proceedings~\citep{camara_statistical_2023}. This section presents the performance of the submitted algorithms, providing an overview of all participating methods included in this benchmark. Publicly available resources (e.g., publications, released code, and Docker images) are summarized in \Cref{tab:overview}.

\subsection{Performance analysis of the IQA task}
\label{sec:quality}
Ten teams successfully completed the final testing phase for the IQA task. An overview of their methodologies is presented in \Cref{tab:task1_overview}. The primary ranking metric for the IQA task was the linearly weighted Cohen's Kappa ($\kappa$), chosen for its robustness in multiclass classification scenarios with potential class imbalance. The final results are reported in \Cref{tab:task1_results}, with overall accuracy provided as a supplementary, non-ranking metric.

\begin{sidewaystable*}[htbp]
  \vspace{9 cm}
  \centering
  \caption{Overview of top 10 methods for IQA task. Type represents the method employ 2D or 3D method. w/ denotes the presence of a given setting, w/o denotes its absence, and n/a indicates that the corresponding result was not reported in the original paper. Ensemble represents the ensemble models during inference time. TTA denotes test time augmentation. ViT represents vision transformer and Swin represents Swin-transformer. SSP represents reconstruction self-supervised pretraining. CNN represents Self-constructed CNN.}
  \resizebox{\linewidth}{!}{
    \renewcommand{\arraystretch}{2.5}
    \begin{tabular}{lcp{8 em}c>{\centering}p{8 em}p{9.125em}p{9.71em}ccp{5.915em}cc}
      \toprule
      \multicolumn{1}{c}{\multirow{2}[4]{*}{Team}} & \multirow{2}[4]{*}{Type} & \multicolumn{2}{c}{Deep Learning Model} & \multicolumn{1}{c}{\multirow{2}[4]{*}{Machine Learning Model}} & \multicolumn{1}{c}{\multirow{2}[4]{*}{Image Preprocessing}} & \multicolumn{1}{c}{\multirow{2}[4]{*}{Data Augmentation}} & \multirow{2}[4]{*}{External Data} & \multirow{2}[4]{*}{Pretraining} & \multicolumn{2}{c}{Ensemble} & \multirow{2}[4]{*}{TTA} \\
  \cmidrule{3-4}\cmidrule{10-11}          &       & CNN Backbone & Transformer Backbone &       & \multicolumn{1}{c}{} & \multicolumn{1}{c}{} &       &       & \multicolumn{1}{c}{Method} & Models &  \\
      \midrule
      A1. UON\_IMA & 2D    & \multicolumn{1}{p{8em}}{ResNet\newline{}EfficientNet} & ViT   & w/o   & Calculatating 2D gradient magnitude map & \multicolumn{1}{c}{n/a} & w/o   & \multicolumn{1}{p{5.915em}}{ImageNet pretraining} & \multicolumn{1}{c}{Majority voting} & 6     & w/o \\
      A2. Philips\_CTS & 3D    & EfficientNet(EF-b0) & w/o   & \multicolumn{1}{p{8em}}{Extra trees, random forest, decision tree, and GDBoost.} & Cropping or padding to 512x512 & Random scale and crop, horizontal flipping, and intensity shift & w/o   & \multicolumn{1}{p{5.915em}}{Simsiam on training set} & \multicolumn{1}{c}{Majority voting} & 5     & w/o \\
      A3. OpenGTN & 2D    & Autoencoder & w/o   & w/o   & K-space motion artifact simulation & \multicolumn{1}{c}{n/a} & \multicolumn{1}{c}{M\&Ms1} & \multicolumn{1}{p{5.915em}}{SSP on training set\newline{} and M\&Ms1} & \multicolumn{1}{c}{n/a} & n/a   & w/o \\
      A4. issun & 2D    & ResNet & w/o   & w/o   & Scale norlization, min-max intensity  normalization & Weighted image interpolation  from same subject, histogram matching with interpolation & w/o   & n/a   & \multicolumn{1}{c}{w/o} & w/o   & w/ \\
      A5. Tewodrosw & 3D    & ResNet-18 & w/o   & w/o   & Resampling, z-score intensity normalization & Random scaling and cropping, random rotation, random flipping  & w/o   & n/a   & \multicolumn{1}{c}{Mean average} & 2     & w/ \\
      A6. CMR.love.LHND & 2D    & \multicolumn{1}{p{8em}}{CNN} & w/o   & w/o   & Image padding & Vertical flip, horizontal flip, and image rotation & w/o   & w/o   & \multicolumn{1}{c}{w/o} & w/o   & w/ \\
      A7. MI-IST\_DA & 2D    & \multicolumn{1}{p{8em}}{ResNet-152\newline{}EfficientNet(EF-b5)} & w/o   & w/o   & Min-max intensity normalization, Center crop & Random affine transformations & w/o   & n/a   & \multicolumn{1}{c}{CORN, CORAL} & 5     & w/o \\
      A8. UA-SVCC & 2D    & EfficientNet(EF-b7) & w/o   & w/o   & Decompose the MR images into 2D slices & Random $90^{\circ}$ rotations, flips, zooms, random Gaussian sharpen and smoothing operations to simulate motion artifacts & w/o   & n/a   & \multicolumn{1}{c}{w/o} & w/o   & w/o \\
      A9. Abdul & 3D    & DenseNet-201 & w/o   & w/o   & Cropping, resampling, z-score intensity normalization & Vertical Flip, Horizontal Flip, Random Gamma,  & w/o   &  w/o   & \multicolumn{1}{c}{w/o} & w/o   & w/o \\
      A10. Sano  & 3D    & w/o   & Swin Transformer & w/o   & z-score intensity normalization & Random flips, random zoom, random rotate & w/o   & n/a   & \multicolumn{1}{c}{5-Fold Ensemble} & 5     & w/o \\
      \bottomrule
      \multicolumn{10}{l}{\textbf{Note}: Deep learning (DL) models refer to architectures based on convolutional neural networks (CNNs), Transformers, and multilayer perceptrons (MLPs), whereas machine learning models refer to classical machine learning approaches such as random forests.} \\
      \multicolumn{10}{l}{SimSiam represents a self-supervised pretraining framework, while CORN and CORAL are learning-based model ensemble methods. 5-Fold in the ensemble column denotes ensembleing models obtained from 5-fold cross-validation.} \\
    \end{tabular}%
  }
  \label{tab:task1_overview}%
\end{sidewaystable*}%
\begin{table*}[htbp]
  \centering
  \caption{Results of the IQA task. Acc represents accuracy and the data representation mode is mean\ \ [$\bold{95\%}$ CI lower, \ $\bold{95\%}$ CI upper]. $\bold{95\%}$ CI represents $\bold{95\%}$ confidence interval. TID represents the team identifier. Values highlighted in bold red denote the best performance, whereas those in bold black indicate the second-best results.}
  \resizebox{\linewidth}{!}{
  \vspace{-2cm}
  \begin{tabular}{>{\centering}p{6em}>{\centering}p{3em}>{\centering}p{10em}>{\centering}p{14em}>{\centering}p{12em}c}
  \toprule
  Rank  & TID & Team & Acc [$\bold{95\%}$ CI] & Cohen's Kappa [$\bold{95\%}$ CI] &  \\
  \midrule
  1     & A1 & UON\_IMA & \textcolor[rgb]{ .753,  0,  0}{\textbf{0.725\ \ [0.642, \ 0.800]}} & \textcolor[rgb]{ .753,  0,  0}{\textbf{0.631\ \ [0.518, \ 0.734]}} &  \\
  2     & A2 & Philips\_CTS & \textbf{0.708\ \ [0.625, \ 0.792]} & \textbf{0.549\ \ [0.419, \ 0.670]} &  \\
  3     & A3 & OpenGTN & 0.625\ \ [0.542, \ 0.708] & 0.474\ \ [0.357, \ 0.582] &  \\
  4     & A4 & issun & 0.642\ \ [0.558, \ 0.725] & 0.456\ \ [0.328, \ 0.580] &  \\
  5     & A5 & Tewodrosw & 0.608\ \ [0.525, \ 0.692] & 0.455\ \ [0.341, \ 0.564] &  \\
  6     & A6 & CMR.love.LHND & 0.650\ \ [0.567, \ 0.733] & 0.447\ \ [0.333, \ 0.561] &  \\
  7     & A7 & MI-IST\_DA & 0.633\ \ [0.550, \ 0.717] & 0.433\ \ [0.325, \ 0.542] &  \\
  8     & A8 & UA-SVCC & 0.608\ \ [0.517, \ 0.692] & 0.432\ \ [0.317, \ 0.540] &  \\
  9     & A9 & Abdul & 0.567\ \ [0.483, \ 0.650] & 0.382\ \ [0.258, \ 0.499] &  \\
  10    & A10 & Sano  & 0.400\ \ [0.317, \ 0.483] & -0.075\ \ [-0.203, \ 0.053] &  \\
  \bottomrule
  \end{tabular}%

  }
  \label{tab:task1_results}%
\end{table*}%

The submitted algorithms demonstrated a wide range of performance on the test set, with Kappa scores ranging from -0.075 to 0.631. The winning team, A1.UON\_IMA, achieved a Kappa score of 0.631 and an accuracy of 0.725. The second and third-place teams, A2.Philips\_CTS and A3.OpenGTN, also delivered strong results, with Kappa scores of 0.549 and 0.474, respectively. A common strategy among the top three teams was the use of pre-training on external datasets, which appeared to confer a significant performance advantage. To assess the statistical significance and stability of these rankings, we performed a bootstrap analysis with 10,000 samples. The resulting 95\% confidence intervals (CI) for the Kappa scores are visualized in \Cref{fig:IQA_Score_CI}. The winning team's score had a tight 95\% CI of [0.518, 0.734], indicating a consistently superior performance. The CIs for the subsequent ranks were wider, with considerable overlap among the teams ranked 4th through 8th, suggesting their performance was statistically similar.

\begin{figure*}[hbt]
  \centering
  \includegraphics[width=\linewidth]{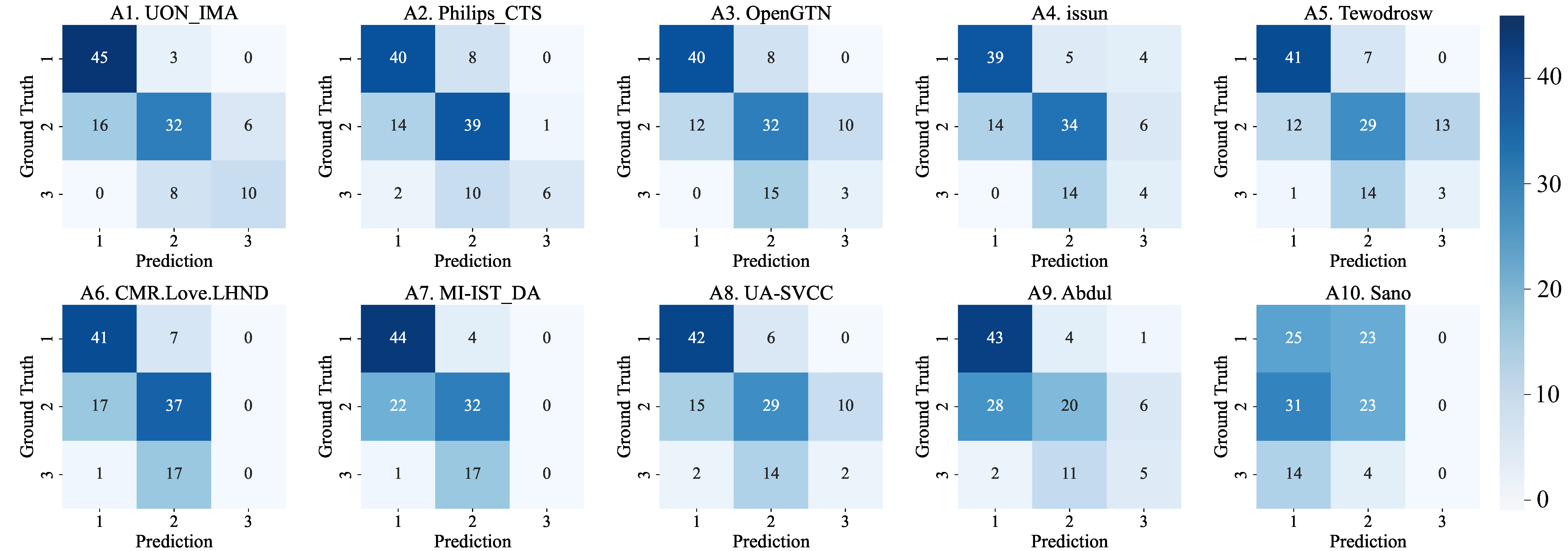}
  \caption{Confusion matrixes for the IQA task. The horizontal axis represents the predicted labels, and the vertical axis represents the ground truth labels.}
  \label{fig:Confusion_mtx}
\end{figure*}

\begin{figure*}[hbtp]
  \centering
  \includegraphics[width=\linewidth]{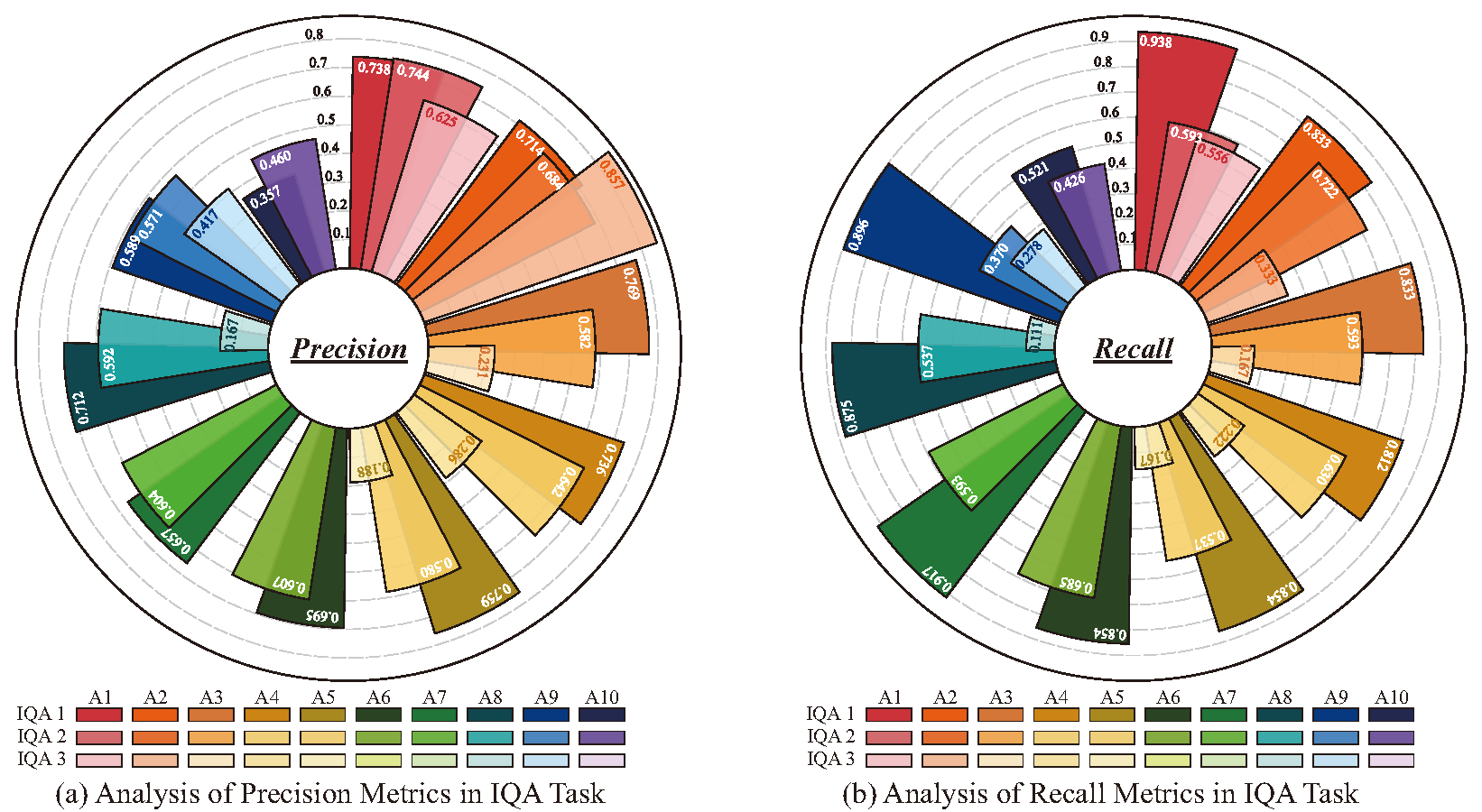}
  \caption{Nightingale rose chart illustrating the per-class classification performance for the IQA task in terms of (a) Precision and (b) Recall. Each colored group corresponds to a participating team, and the three sectors within each group represent performance on the three IQA labels.}
  \label{fig:rose_chart}
\end{figure*}

A detailed breakdown of per-class performance is provided by the confusion matrixes in \Cref{fig:Confusion_mtx} and the precision-recall analysis in \Cref{fig:rose_chart}. These results highlight a key challenge: while most algorithms performed well in identifying images with mild motion (Label 1), their performance degraded significantly for images with intermediate (Label 2) and, particularly, severe motion (Label 3). As shown in \Cref{fig:rose_chart} (b), the recall for severe motion artifacts was generally low across all teams, indicating that these cases were the most difficult to correctly classify. This suggests that while current methods are effective at identifying high-quality images, accurately classifying the degree of severe motion artifacts remains a significant area for future improvement.

\subsection{Overview of IQA methodologies}
The ten submitted solutions for the IQA task were all based on machine learning, with a strong predominance of deep learning techniques. A summary of the methods and their configurations is provided in \Cref{tab:task1_overview}.

\subsubsection{Architectures and training paradigms}
All submitted algorithms predominantly employed deep learning (DL) methods, involving convolutional neural networks (CNNs) or Transformer architectures. Notably, one team~\citep{li-automatic-2022} integrated DL techniques with classical machine learning models (including Extra Trees, Random Forest, Decision Tree, and GDBoost classifiers) by leveraging radiomics-derived features. The adopted end-to-end learning frameworks incorporated many state-of-the-art (SOTA) classification models. Popular choices of CNN backbones included the ResNet~\citep{He_Res_2016} and EfficientNet~\citep{2019arXiv190511946T} families. Transformer-based models included Vision Transformers (ViT) and Swin Transformers. As shown in \Cref{tab:task1_results}, it worth to note a notable trend that lightweight backbones~\citep{li-automatic-2022, arega-automatic-2022, yang-deep-2022} often achieved competitive or even superior performance, possibly due to a reduced risk of overfitting on the limited dataset. The majority of participating teams adopted 2D models that processed CMR volumes in a slice-wise manner, which generally outperformed their 3D counterparts.

Most teams utilized a standard, end-to-end supervised training framework. However, several explored alternative paradigms to boost performance. Multi-task learning was adopted by some, either by treating IQA as the primary task with an auxiliary objective~\citep{arega-automatic-2022} or vice versa~\citep{grzeszczyk_multi-task_2022}. Pre-training emerged as a particularly effective strategy, with several top teams leveraging transfer learning from models pre-trained on ImageNet~\citep{li-motion-related-2022}. Self-supervised pre-training (SSP) was also successfully employed, using techniques like SimSiam~\citep{Chen_2021_CVPR} to enhance representation learning~\citep{li-automatic-2022} or by training an autoencoder to reconstruct clean images from synthetically corrupted ones with simulated motion artifacts~\citep{amirrajab-cardiac-2022}.

\subsubsection{Ensemble strategies}
Model ensembling was a key strategy for many of the top-performing teams. These approaches varied in complexity and design, encompassing architectural ensembles and data-level ensembles (e.g., combining models at the slice level and volume level). The winning team, UON\_IMA, developed a hierarchical ensemble that first combining predictions from different 2D CNN and ViT models at the slice level, and then aggregating these into a final volume-level prediction using a custom voting scheme~\citep{li-motion-related-2022}. Another successful approach, proposed by Philips\_CTS through the RE-Vote framework, combined a deep learning model with four classical machine learning classifiers trained on radiomic features~\citep{li-automatic-2022}. Learning-based ensembles, such as rank-consistent networks like CORAL~\citep{cao_rank_2020} and CORN~\citep{10.1007/s10044-023-01181-9} also exhibited performance advantages over individual models. Simpler but still effective ensemble methods included combining models trained with different loss functions~\citep{arega-automatic-2022} or aggregating predictions from models trained across different cross-validation folds~\citep{mora-rubio-deep-2022}.

\subsubsection{Data preprocessing and augmentation}
Nearly all teams employed a series of preprocessing steps to standardize the input data. Common techniques included resampling to a uniform voxel spacing, as well as cropping and padding to a fixed volume size. For intensity normalization, both z-score standardization and min-max scaling were widely used. Some teams also incorporated more specialized preprocessing, such as generating 2D gradient maps as an additional input modality~\citep{li-motion-related-2022} or simulating motion artifacts in k-space~\citep{amirrajab-cardiac-2022}.

Data augmentation was universally applied to increase the diversity of the training set and improve model robustness. Standard spatial augmentations included random rotations, scaling, cropping, and flips. Intensity-based augmentations involved adjustments to brightness and contrast, as well as the application of Gaussian blurring and sharpening to simulate motion-like effects. Several teams also employed Test-Time Augmentation (TTA), where multiple augmented versions of a test image are evaluated and their predictions averaged to produce a more robust final output~\citep{sun_combination_2022, arega-automatic-2022, yang-deep-2022}.

\begin{figure*}[hbt]
  \centering
  \includegraphics[width=\linewidth]{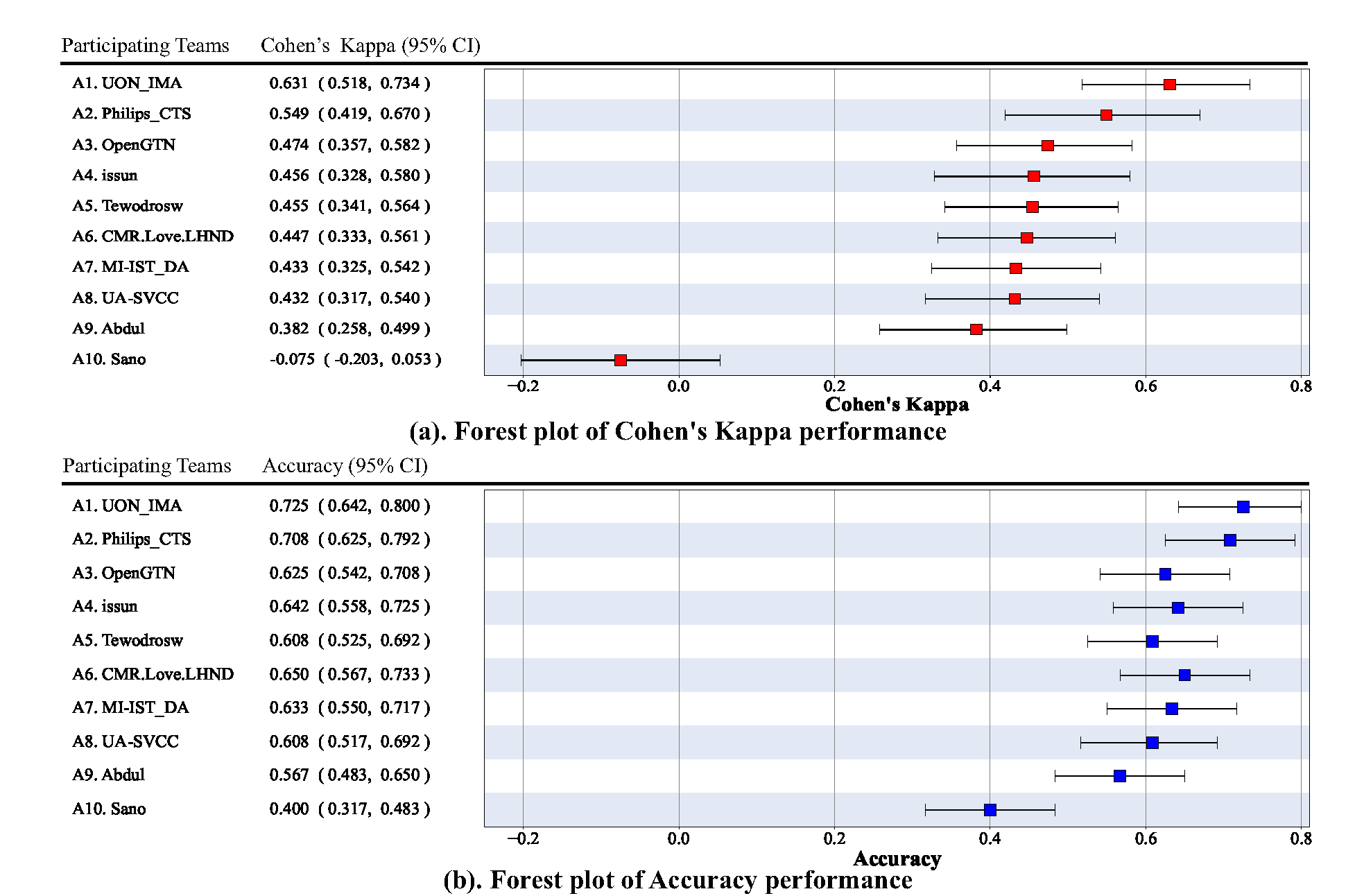}
  \caption{
    Forest plots of IQA performance. The means (depicted as red and blue squares) and the $\bold{95\%}$ confidence intervals (shown as horizontal lines) are estimated from $\bold{10,000}$ bootstrap samples of the image quality assessment task on the test dataset. The x-axis represents the metric values, while the y-axis denotes the participating teams. The horizontal lines indicate the $\bold{95\%}$ confidence intervals ($\bold{95\%}$ CI) derived from the bootstrap analysis. The means (represented by red and blue squares) are calculated by averaging the respective metric across all cases of test set.}
  \label{fig:IQA_Score_CI}
\end{figure*}

\subsection{Details of IQA algorithms}
\label{sec:iqa_sota}
\subsubsection{A1. UON\_IMA}
The team UON\_IMA~\citep{li-motion-related-2022} won first place in the IQA task. They developed an ensemble learning framework incorporating several 2D advanced classification models (i.e., ResNet~\citep{He_Res_2016}, EfficientNet~\citep{2019arXiv190511946T}, and ViT~\citep{dosovitskiy_image_2020}) pretrained on ImageNe~\citep{5206848} on both the original intensity images and the 2D in-plane gradient magnitude maps. Observations indicated that slices from the apical and basal regions may contribute minimally to the overall IQA score. To address this, the method first employs a U-Net~\citep{ronneberger_u-net_2015} to locate and sample image patches around the foreground region. Six classification models are then applied, derived by training each of the three base architectures (ResNet, EfficientNet, ViT) separately on both image modalities (intensity and gradient). The final predictions are integrated using a hierarchical voting strategy: after obtaining scores for each slice, a subject-level decision is made for each model using a bias voting scheme on the slice-level predictions. These subject-level predictions are then combined via a final majority vote. The bias voting scheme incorporates weighting factors derived from 5-fold cross-validation on the training set. The proposed ensemble method demonstrates robustness and efficiency on both the validation and testing datasets, outperforming other approaches as presented in \Cref{tab:task1_results} and \Cref{fig:IQA_Score_CI}. 

\subsubsection{A2. Philips\_CTS}
The team Philips\_CTS~\citep{li-automatic-2022} also employed an ensemble framework using both DL models and machine learning approaches based on radiomics, achieving second place in the IQA task. The deep-learning model employed is a lightweight EfficientNet-B0~\citep{2019arXiv190511946T}. To mitigate overfitting due to the limited dataset size, self-supervised pretraining— specifically SimSiam~\citep{Chen_2021_CVPR}—was utilized to enhance representation learning and alleviate overfitting in EfficientNet-B0. Subsequently, the authors applied four machine learning models (Extra Trees, Random Forest, Decision Tree, and Gradient Boosting (GDBoost) classifiers) to generate radiomics-based predictions based on 50 selected features, which were extracted and selected via PyCaret~\citep{PyCaret} from the LV regions identified and cropped using nnU-Net~\citep{isensee_nnu-net_2021}. By employing a weighted voting mechanism — assigning weights of 0.2 to deep learning and 0.8 to the radiomics models— Philips\_CTS achieved a Cohen's Kappa score of 0.549, with a $95\%$ CI of $[0.419, 0.670]$.

\subsubsection{A3. OpenGTN}
The OpenGTN~\citep{amirrajab-cardiac-2022} team secured third place by training an autoencoder—a 2D U-Net-like model without skip connections—to take input images with simulated motion artifacts and reconstruct the original artifact-free images. Following this motion-denoising self-supervised pretraining, the encoder was connected to two fully connected layers to predict the IQA score for each slice. The final IQA score for the entire image volume was obtained by averaging the slice-level scores and applying a conditional decision rule. To augment the training dataset, k-space motion artifact simulation was used to generate SSP training data. This was achieved by introducing simulated artifacts into motion-free images from the \textit{CMRxMotion} training dataset, as well as incorporating additional data from the M\&Ms-1 challenge~\citep{9458279}. With this approach and the implementation of targeted data augmentation strategies, OpenGTN achieved a Cohen's Kappa score of 0.474, with a $95\%$ CI of $[0.357, 0.582]$.

\subsubsection{A4. issun}
Team issun~\citep{sun_combination_2022} introduced a two-branch sampling inspection network with dedicated data augmentation techniques for IQA task. The network emulates human visual inspection by applying two distinct random slice sampling strategies—one per branch. Each branch employs ResNet as its backbone to extract features, which are then fused through a dedicated fusion module to generate the final IQA score.

\subsubsection{A5. Tewodrosw}
Team Tewodrosw~\citep{arega-automatic-2022} developed a multi-task learning framework based on 3D ResNet-18~\citep{He_Res_2016}, incorporating a second classification branch to predict the patient's breath-hold type as an auxiliary task. This multi-task setup demonstrated significant improvements over single-task IQA prediction. To further enhance prediction accuracy, the method applies TTA techniques, averaging predictions over $M$ augmented samples to produce the final output. Additionally, an ensemble of models—comprising both single-task and multi-task variants trained with different loss functions (focal loss and weighted cross-entropy)—provided an additional performance gain. With this comprehensive strategy, Team Tewodrosw secured $5_{th}$ place in the benchmark.

\subsubsection{A6. CMR.Love.LHND}
Subsequently ranked in the benchmark is Team CMR.Love.LHND~\citep{yang-deep-2022}, which treated the IQA task as a quality control problem and developed a lightweight 2D CNN model. By incorporating basic techniques such as residual connections, the team achieved a balance between strong performance and memory efficiency, with the model requiring only 3 GB of video memory. To further enhance performance, TTA techniques was employed, averaging predictions across three image transformations to produce more robust motion-level estimations. Notably, while this method achieved the highest score on the live validation leaderboard, it ultimately placed sixth on the final test set.

\subsubsection{A7. MI-IST\_DA}
Team Aranme~\citep{ranem_detecting_2022} introduced the rank-consistent neural network, namely CORAL~\citep{cao_rank_2020} and CORN~\citep{10.1007/s10044-023-01181-9}, to make the final decisions using several binary classifiers while taking into account the rank consistency among predictions. By using CORAL and adopting EfficientNet-B5 as the backbone, a significant improvement over ResNet-152 is observed in the IQA task. This method secured the $6_{th}$ place in our benchmark.

\subsubsection{A8. UA-SVCC}
Team UA-SVCC~\citep{mora-rubio-deep-2022} employed EfficientNet-b7~\citep{2019arXiv190511946T} for the IQA task. To train this large-scale backbone, the team used a weighted loss function along with the Nadam optimizer~\citep{dozat2015technical}. The CMR volumes were decomposed into 2D slices, and a variety of augmentation techniques were applied, including random $90^{\circ}$ rotations, flips, and zooms. Additionally, random Gaussian sharpening and smoothing operations were introduced to simulate the effects of respiratory motion artifacts.

\subsubsection{A9. Abdual}
Team Abdul~\citep{qayyum_automatic_2022} introduced a 3D DenseNet-201 architecture comprising four dense blocks, each containing six layers, based on the work of \citep{Huang_2017_CVPR}. This design leverages dense connectivity to address the vanishing gradient problem, enhance feature propagation, encourage feature reuse, and reduce the overall number of parameters. In the proposed framework, only a small subset of features is added at each layer, while the remaining feature maps are preserved and reused throughout the network. The final decision layer utilizes all accumulated feature maps, enabling comprehensive representation learning. This approach not only reduces parameter count but also improves the flow of information and gradients, simplifying the training process. Importantly, each layer has direct access to the loss gradients and input data, enforcing an implicit deep supervision mechanism that facilitates training of deeper networks. Additionally, the dense connections exhibit a regularizing effect, which helps mitigate overfitting even with limited training data.

\begin{sidewaystable*}
  \vspace{9 cm}
  \centering
  \caption{Overview of RCS methods. Type represents the method employ 2D or 3D method. w/ denotes the presence of a given setting, w/o denotes its absence, and n/a indicates that the corresponding result was not reported in the original paper. Ensemble represents the ensemble models during inference time. TTA denotes test time augmentation. CE represent the cross entropy loss, and HD represent loss function based on Hausdorff Distance.}
  \resizebox{\linewidth}{!}{
  \renewcommand{\arraystretch}{2.5}
  \begin{tabular}{llllp{10em}lp{14em}llllp{8em}lll}
    \toprule
    \multicolumn{1}{l}{\multirow{2}[4]{*}{Team}} & \multicolumn{1}{c}{\multirow{2}[4]{*}{Type}} & \multicolumn{1}{c}{\multirow{2}[4]{*}{CNN Backbone}} & \multicolumn{1}{c}{\multirow{2}[4]{*}{Transformer backbone}} & \multirow{2}[4]{*}{Preprocessing} & \multicolumn{1}{c}{\multirow{2}[4]{*}{Post-processing}} & \multirow{2}[4]{*}{Data Augumentation} & \multicolumn{1}{c}{\multirow{2}[4]{*}{Pesudo Label}} & \multicolumn{1}{c}{\multirow{2}[4]{*}{Externel Data}} & \multicolumn{1}{c}{\multirow{2}[4]{*}{Pretrained Model}} & \multicolumn{2}{p{16.79em}}{Uncertainty Estimation Methods} & \multicolumn{1}{c}{\multirow{2}[4]{*}{Multi-stage}} & \multicolumn{1}{c}{\multirow{2}[4]{*}{Training}} & \multicolumn{1}{c}{\multirow{2}[4]{*}{Loss Function}} \\
\cmidrule{11-12}          &       &       &       & \multicolumn{1}{c}{} &       & \multicolumn{1}{c}{} &       &       &       & \multicolumn{1}{p{7.54em}}{TTA} & \multicolumn{1}{p{9.25em}}{Ensemble} &       &       &  \\
    \midrule
    S1. UA-SVCC & 2D    & nnU-Net & w/o   & nnU-Net  preprocessing & SPP     & Random 90 rotations, flips, zooms, random Gaussian sharpen and smoothing & w/o   & w/o   & n/a   & n/a   & n/a   & w/  & FSL   & Dice+CE \\
    \multirow{3}[16]{*}{S2. Med-Air} & 2D, 3D & nnU-Net & w/o   & nnU-Net  preprocessing & \multirow{3}[0]{*}{SPP} & \multirow{3}[0]{*}{\parbox[c]{14em}{Random rotation, scaling, Gaussian noise, Gaussian blur, brightness, contrast, simulation of low resolution, gamma correction, and mirroring,  adversarial augmentation (AdvChain)}} & \multirow{3}[0]{*}{w/o} & \multicolumn{1}{r}{\multirow{3}[0]{*}{\parbox[c]{8em}{SCD, LASB'13, RVSC, DSB ACDC, M\&Ms-1, M\&Ms-2, MyoPS, MS-CMRSeg}}} & \multicolumn{1}{p{11.875em}}{PT on ACDC, M\&Ms-1, M\&Ms-2, MyoPS, MS-CMRSeg} & \multirow{3}[0]{*}{w/} & \multicolumn{1}{l}{\multirow{3}[16]{*}{\parbox[c]{8em}{5-Fold  Stratified Cross-validation}}} & \multirow{3}[16]{*}{w/o} & \multirow{3}[16]{*}{FSL}   & \multirow{3}[16]{*}{Dice+CE} \\
          & 2D    & w/o   & \multicolumn{1}{p{8.085em}}{Swin-UNETR} & \multirow{2}[0]{*}{\parbox[c]{10em}{Center Cropping, Intensity normilization\newline{} (min-max)}} &       & \multicolumn{1}{r}{} &       &       & \multicolumn{1}{p{11.875em}}{SSL on SCD, LASB'13, RVSC, DSB ACDC and PT on ACDC, M\&Ms-1, M\&Ms-2, MyoPS, MS-CMRSeg} &       &       &       &       &  \\
          & 2D    & w/o   & \multicolumn{1}{p{8.085em}}{Swin-UNet} & \multicolumn{1}{r}{} &       & \multicolumn{1}{r}{} &       &       & \multicolumn{1}{p{11.875em}}{SSL on ImageNet and PT on ACDC, M\&Ms-1, M\&Ms-2, MyoPS, MS-CMRSeg} &       &       &       &       &  \\
    \multirow{2}[0]{*}{S3. CMR.Love.LHND} & 2D    & \multicolumn{1}{p{8.835em}}{Bi-direaction LSTM} & \multicolumn{1}{l}{\multirow{2}[0]{*}{w/o}} & w/o   & w/o   & \multirow{2}[0]{*}{\parbox[c]{14em}{enhanced Gaussian noise, Gaussian blur, brightness, contrast, and gamma correction}} & \multirow{2}[0]{*}{w/o} & \multirow{2}[0]{*}{w/o} & \multirow{2}[0]{*}{w/o} & \multirow{2}[0]{*}{w/o} & \multicolumn{1}{l}{\multirow{2}[0]{*}{\parbox[c]{8em}{Ensemble of a 2D and a 3D version}}} & \multirow{2}[0]{*}{w/o} & \multirow{2}[0]{*}{FSL}   & \multirow{2}[0]{*}{Dice+CE+HD} \\
          & 2D, 3D & \multicolumn{1}{p{8.835em}}{nnU-Net} &       & nnU-Net  preprocessing & nnU-Net  postprocessing & \multicolumn{1}{r}{} &       &       &       &       &       &       &       &  \\
    S4. Tewodrosw & 3D    & nnU-Net & w/o   & Image resampling & n/a   & k-space augumentation, brightness, contrast, gamma, Gaussian noise,  blur. Elastic deformation, random rotation, random scaling, random flipping, and low spatial resolution simulation. & w/o   & \multicolumn{1}{p{8.75em}}{ACDC} & n/a   & w/o   & w/o   & w/o   & FSL   & DicePolyCE \\
    S5. OpenGTN & 2D    & nnU-Net & w/o   & nnU-Net  preprocessing & n/a   & Augmentation with simulated motion artifacts & w/o   & \multicolumn{1}{p{8.75em}}{M\&Ms-1,  M\&Ms-2 } & n/a   & n/a   & \multicolumn{1}{p{9.25em}}{Ensemble at the instance and slice Levels} & w/     & FSL   & \multicolumn{1}{p{9.46em}}{Focal-Tversky\newline{}Dice+CE} \\
    S6. Abdul & 3D    & \multicolumn{1}{p{8.835em}}{nn-UNet\newline{}3D-ResUnet} & w/o   & Resampling (nearest neighbor interpolation), Intensity Normalization (z-score) & \multicolumn{1}{p{11em}}{Volume resizing (bilinear interpolation)} & Horizontal Flip, Vertical Flip, Random Gamma & w/     & w/o   & n/a   & w/o   & Cross-validation & Two stage & FSL+SSL & Dice+BCE \\
    S7. Philips\_CTS & 2D    & nnU-Net & w/o   & Spatial normalization, intensity normalization (min-max) & n/a   & Rotation, flipping, scaling, deformation of the original images, adding noise,  brigh tness modification & w/     & w/o   & n/a   & n/a   & w/o   & Two stage & FSL+SSL & Dice+CE \\
    \multirow{2}[0]{*}{S8. MI-IST\_DA} & 2D    &         & ViT U-Net V2 & \multirow{2}[0]{*}{nnU-Net preprocessing} & \multirow{2}[0]{*}{nnU-Net  postprocessing} & \multirow{2}[0]{5cm}{Augmentation with simulated motion artifacts, flipping, rotation} & w/o   & w/o   & w/o   & w/o   & \multirow{2}[0]{*}{\parbox[c]{10em}{Ensemble of ViT U-Net V2, and nnU-Net (2D amd 3D)}} & \multirow{2}[0]{*}{w/o} & \multirow{2}[0]{*}{FSL} &  \multirow{2}[0]{*}{Dice+CE} \\
                                       & 2D,3D & nnU-Net &              &                                           &                                             &                                                                                      & w/o   & w/o   & n/a   & w/o   &                         &                         &                         &                              \\
    S9. ML-Labs & 2D    & \multicolumn{1}{p{8.835em}}{U-Net (ResNet101)} & w/o   & Image resampling, Image orientation normalization,  ROI cropping, Intensity normalization (histogram standardization) & w/o   & Random Motion,Random Ghosting,Random Bias Field,Random Gamma & w/o   & w/o   & PT on ImageNet & n/a   & n/a   & w/o   & FSL   & Dice+CE \\
    S10. Sano  & 3D    & w/o   & \multicolumn{1}{p{8.085em}}{Swin UNETR} & Volume resizing,volume padding & n/a   & Random flips, random zoom, random rotation & w/o   & w/o   & w/o   & w/o   & \multicolumn{1}{p{9.25em}}{5-Fold  Stratified Cross-validation} & w/o   & FSL   & Dice+CE \\
    S11. HAHA2 & 2D    & \multicolumn{1}{p{8.835em}}{DeepLabv3+\newline{}ResNet101} & w/o   & n/a   & n/a   & Fourier transformation & w/     & w/o   & w/     & w/o   & w/     & w/o   & FSL   & Dice+CE \\
    S12. sots  & 3D    & DenseBiasNet+VAE & w/o   &  Intensity normalization (z-score) & n/a   & n/a   & w/o   & w/o   & n/a   & w/o   & n/a   & w/o   & FSL   & SoftDice+CE+L2 \\
    \bottomrule
    \multicolumn{10}{l}{DSB: Second Annual Data Science Bowl. LASC'13: Left Atrial Segmentation Challenge 2013. RVSC Right Ventricle Segmentation Challenge (RVSC).} \\
    \multicolumn{10}{l}{SCD: Sunnybrook Cardiac Data (SCD), also known as the 2009 Cardiac MR Left Ventricle Segmentation Challenge data.} \\
    \multicolumn{10}{l}{PT: Pre-training, SSL: Self-Supervised Learning. SPP: Specilized Post-processing.} \\
    \end{tabular}%
  }
  \label{tab:task2_overview}%
\end{sidewaystable*}%

\subsubsection{A10. Sano}
Team Sano~\citep{grzeszczyk_multi-task_2022} adopted a multi-task learning framework for simultaneous IQA and segmentation, based on the Swin-UNETR architecture~\citep{hatamizadeh_swin_2022}. In this design, a 3D Swin Transformer encoder, originally used for segmentation, is connected to additional fully connected layers following global average pooling and dropout. This auxiliary branch is responsible for predicting the IQA score in parallel with the segmentation task. The final IQA score is obtained by ensembling predictions from models trained using a 5-fold cross-validation strategy. However, this approach did not yield a significant improvement in kappa score, potentially due to the substantial task differences between segmentation and IQA, or as a result of overfitting on the validation or local dataset.

\subsection{Performance analysis of the RCS task}
\label{sec:segmentation}

\begin{figure*}[htbp]
  \centering
  \includegraphics[width=\linewidth]{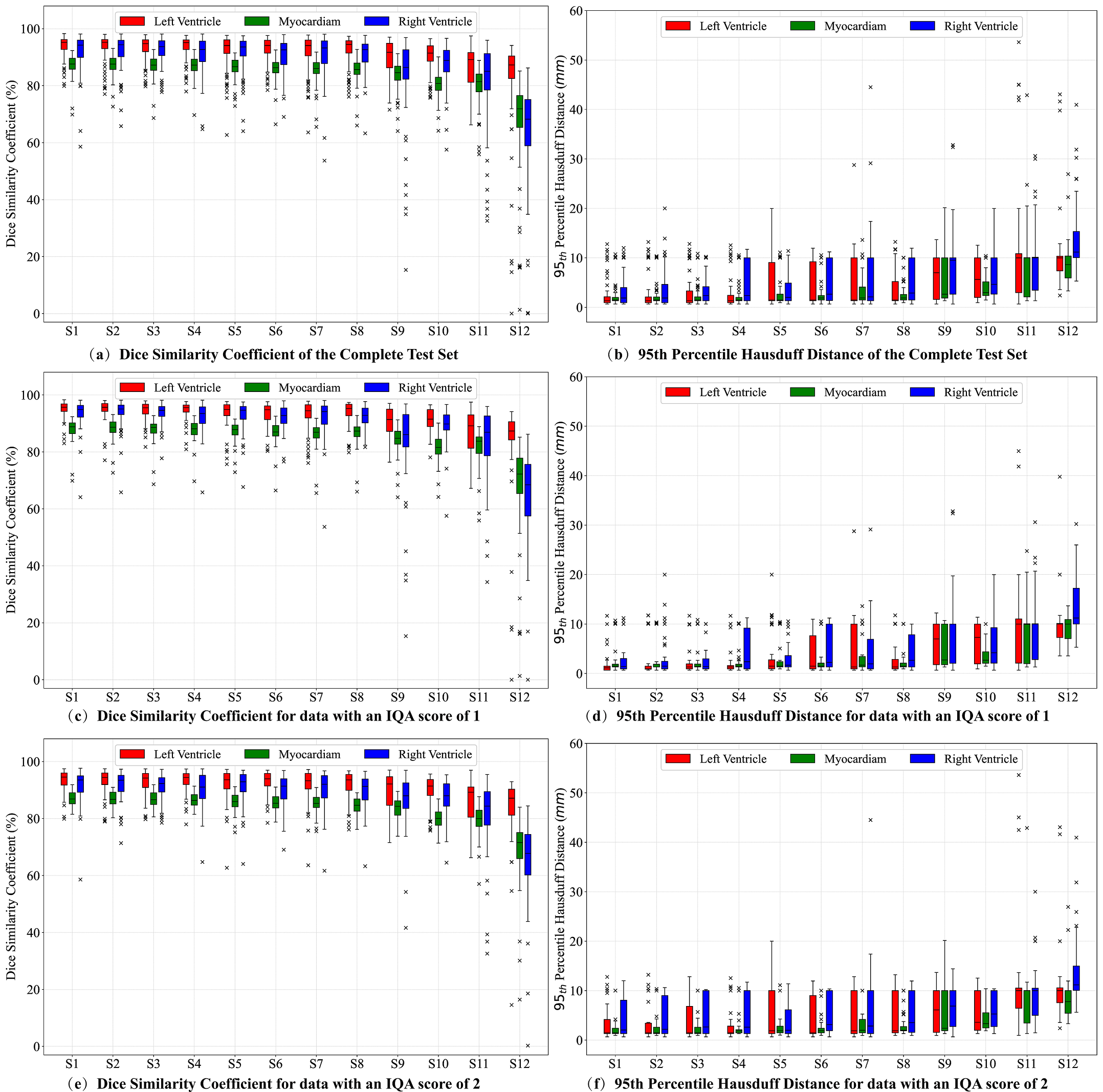}
  \caption{Box plots of the quantitative results for the robust segmentation task on the test set. The plots show the distribution of (a, c, e) Dice Similarity Coefficient and (b, d, f) 95th percentile Hausdorff Distance across the full test set and subsets with different IQA scores. The midline represents the median, and outliers are marked with $\times$.}
  \label{fig:box_plot}
\end{figure*}

\begin{table*}[hbtp]
      \centering
      \caption{Performance of the RCS Task (Task 2). The reported data are presented in the format of mean $\pm$ standard deviation. $\uparrow$ indicates that higher values represent better performance, while a downward arrow ($\downarrow$ ) indicates that lower values are preferred. Values highlighted in bold red denote the best performance, whereas those in bold black indicate the second-best results. LV represents left ventricle, MYO represents myocardium, and RV representation right ventricle respectively.}
      \resizebox{\linewidth}{!}{
        \begin{tabular}{c>{\raggedright\arraybackslash}p{9.2em}*{3}{>{\centering}p{6.5em}}c*{1}{>{\centering}p{6.5em}}cp{6.5em}}
        \toprule
        \multicolumn{1}{c}{\multirow{2}[4]{*}{\textbf{Rank}}} & \multicolumn{1}{c}{\multirow{2}[4]{*}{\textbf{Team}}} & \multicolumn{3}{c}{\textbf{DSC (\%)} $\uparrow$}  &       & \multicolumn{3}{c}{\textbf{HD95 ($mm$)} $\downarrow$} \\
        \cmidrule{3-5}\cmidrule{7-9}          & \multicolumn{1}{c}{} & \textbf{LV} & \textbf{MYO} & \textbf{RV} &       & \textbf{LV} & \textbf{MYO} & \textbf{RV} \\
        \midrule
        \multirow{2}[1]{*}{1} & S1. UA-SVCC & \textcolor[rgb]{ .753,  0,  0}{\textbf{93.90 $\pm$ 3.72}} & \textbf{87.37 $\pm$ 3.36} & 92.22 $\pm$ 6.07 &       & \textbf{3.05 $\pm$ 3.54} & \textbf{2.20 $\pm$ 2.10} & \textcolor[rgb]{ .753,  0,  0}{\textbf{3.53 $\pm$ 3.55}} \\
              & S2. Med-Air & 93.72 $\pm$ 4.32 & \textcolor[rgb]{ .753,  0,  0}{\textbf{87.38 $\pm$ 3.25}} & \textcolor[rgb]{ .753,  0,  0}{\textbf{92.39 $\pm$ 5.52}} & \textcolor[rgb]{ .753,  0,  0}{} & 3.13 $\pm$ 3.71 & 2.57 $\pm$ 2.70 & 3.59 $\pm$ 3.85 \\
        \multirow{2}[0]{*}{2} & S3. CMR.Love.LHND   & 93.41 $\pm$ 4.10 & 87.06 $\pm$ 3.56 & \textbf{92.34 $\pm$ 4.40} &       & 3.42 $\pm$ 3.74 & 2.37 $\pm$ 2.26 & 3.68 $\pm$ 3.40 \\
              & S4. Tewodrosw & \textbf{93.80 $\pm$ 3.77} & 87.08 $\pm$ 3.20 & 91.14 $\pm$ 5.93 &       & \textcolor[rgb]{ .753,  0,  0}{\textbf{2.95 $\pm$ 3.39}} & \textcolor[rgb]{ .753,  0,  0}{\textbf{2.11 $\pm$ 1.78}} & 4.68 $\pm$ 4.03 \\
        3     & S5. OpenGTN & 92.45 $\pm$ 5.67 & 86.32 $\pm$ 3.51 & 91.82 $\pm$ 5.89 &       & 3.98 $\pm$ 4.44 & 3.16 $\pm$ 3.17 & \textbf{3.54 $\pm$ 3.36} \\
        4     & S6. Abdul & 92.82 $\pm$ 4.27 & 86.13 $\pm$ 3.54 & 90.67 $\pm$ 5.57 &       & 3.89 $\pm$ 3.83 & 2.61 $\pm$ 2.39 & 4.87 $\pm$ 3.90 \\
        5     & S7. Philips\_CTS & 91.90 $\pm$ 6.15 & 85.54 $\pm$ 4.09 & 90.75 $\pm$ 7.15 &       & 4.92 $\pm$ 7.47 & 3.66 $\pm$ 3.41 & 5.13 $\pm$ 6.20 \\
        6     & S8. MI-IST\_DA & 92.53 $\pm$ 5.04 & 85.55 $\pm$ 3.93 & 90.80 $\pm$ 5.40 &       & 3.76 $\pm$ 3.85 & 2.76 $\pm$ 2.48 & 4.87 $\pm$ 3.71 \\
        7     & S9. ML-Labs & 89.65 $\pm$ 6.29 & 83.69 $\pm$ 4.55 & 84.22 $\pm$ 13.68 &       & 6.17 $\pm$ 4.22 & 5.03 $\pm$ 4.06 & 7.69 $\pm$ 5.98 \\
        8     & S10. Sano  & 90.31 $\pm$ 5.01 & 80.53 $\pm$ 4.21 & 88.07 $\pm$ 6.27 &       & 6.04 $\pm$ 3.96 & 4.12 $\pm$ 2.56 & 5.75 $\pm$ 3.64 \\
        9     & S11. HAHA2  & 86.59 $\pm$ 7.52 & 80.25 $\pm$ 6.18 & 81.51 $\pm$ 14.00 &       & 12.26 $\pm$ 14.91 & 7.63 $\pm$ 5.69 & 9.28 $\pm$ 6.42 \\
        10    & S12. sots  & 81.81 $\pm$ 18.31 & 67.73 $\pm$ 16.05 & 63.95 $\pm$ 17.31 &       & 15.21 $\pm$ 21.48 & 17.53 $\pm$ 28.91 & 21.16 $\pm$ 50.95 \\
        \bottomrule
        \end{tabular}%
      }
      \label{tab:task2_results}%
\end{table*}%

Thirteen teams successfully completed the final testing phase for the RCS task. This section provides a detailed analysis of the submitted algorithms' performance, followed by an overview of the common methodological themes that emerged from the top-performing solutions. An overview of the specific methods is provided in \Cref{tab:task2_overview}.

The quantitative results for all participating teams are summarized in \Cref{tab:task2_results}. The submitted algorithms achieved a wide range of segmentation accuracies. The average Dice Similarity Coefficient (DSC) scores across all teams ranged from 81.81\% to 93.90\% for the LV, 67.73\% to 87.37\% for the MYO, and 63.95\% to 92.22\% for the RV. The corresponding average 95\% Hausdorff Distance (HD95) values ranged from 3.05 mm to 15.21 mm for the LV, 2.20 mm to 17.53 mm for the MYO, and 3.53 mm to 21.16 mm for the RV.

Consistent with previous cardiac segmentation challenges, the MYO proved to be the most challenging structure to segment, primarily due to its thin anatomical structure. In contrast, the LV and RV blood pools were generally segmented with higher accuracy. In terms of the DSC metric, the top performance for LV segmentation was achieved by UA-SVCC (93.90\%), while Med-Air obtained the best scores for the MYO (87.38\%) and RV (92.39\%). For the HD95 metric, Tewodrosw achieved the lowest HD95 distance for the LV (2.95 mm) and MYO (2.11 mm), while UA-SVCC obtained the best performance for the RV (3.53 mm).

\begin{figure*}[hbpt]
  \centering
  \includegraphics[width=\linewidth]{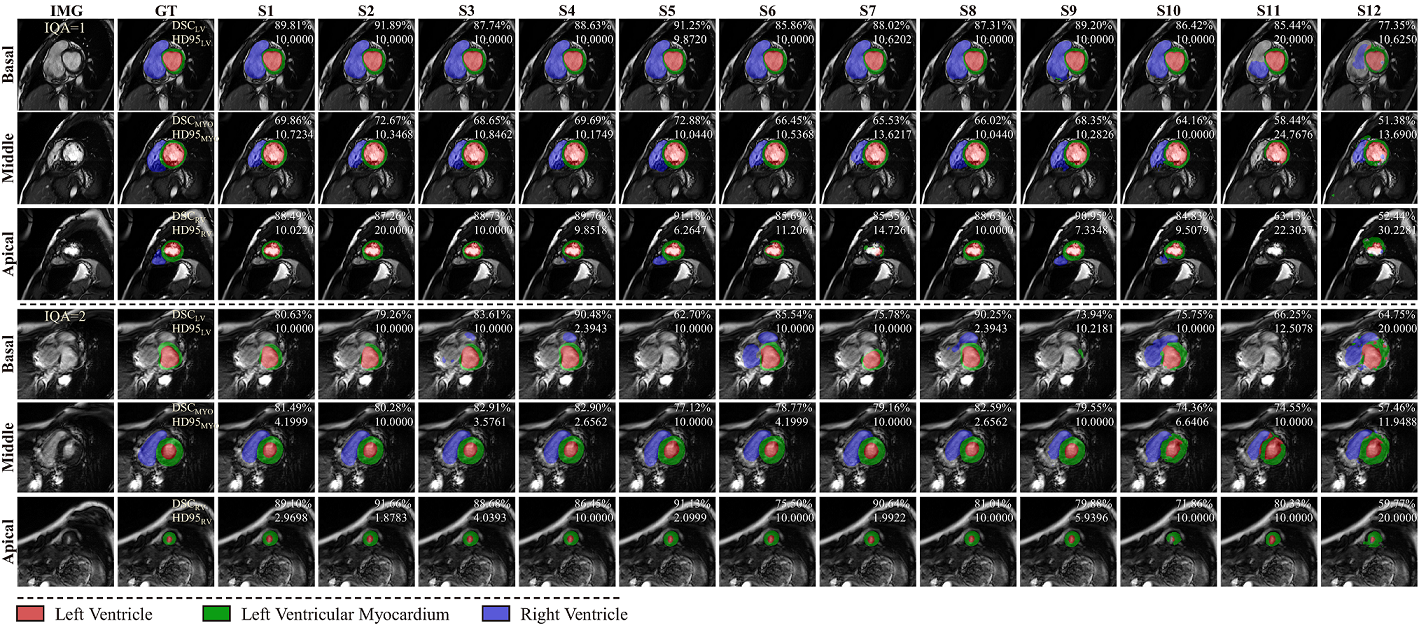}
  \caption{Qualitative segmentation results from all participants for two challenging cases with IQA scores of 1 (top three rows) and 2 (bottom three rows). For each case, the basal, middle, and apical slices are shown. GT denotes the ground truth.}
  \label{fig:SegResults}
\end{figure*}

The final ranking was determined using a robust rank-then-aggregate strategy, with the Wilcoxon signed-rank test used to assess statistical significance. This analysis resulted in a tie for first place between S1.UA-SVCC and S2.Med-Air, and a tie for second place between S3.CMR.Love.LHND and S4.Tewodrosw. The box plots in \Cref{fig:box_plot} illustrate the distribution of scores, showing that while the top teams achieved similar median performances, the winning methods exhibited greater consistency with tighter distributions and fewer outliers. The qualitative results in \Cref{fig:SegResults} further highlight that segmentation of the basal and apical slices was particularly challenging across all methods.

\subsection{Overview of RCS methodologies}
The submitted solutions for the RCS task were dominated by deep learning-based approaches, with several common themes emerging in terms of network architecture, training strategies, and data handling.

\subsubsection{Architectures and training paradigms}
For RCS task, the vast majority of submissions were based on the encoder-decoder architecture like U-Net~\citep{ronneberger_u-net_2015} or the U-Net varients~\citep{qayyum_automatic_2022, garcia-cabrera_cardiac_2022}, with the nnU-Net framework~\citep{nnUNet} being particularly popular. Seven of the twelve teams employed nnU-Net either as their primary model, as a key component within an ensemble, or as a critical stage in a multi-step pipeline. Other architectures included Transformer-based models, such as Swin-UNETR~\citep{hatamizadeh_swin_2022} and ViT U-Net~\citep{Ranem_2022_CVPR}. Although these generally did not outperform the top CNN-based approaches, particularly for the fine-structured MYO, they showed promising results in segmenting larger, convex-shaped structures such as the LV and RV. As shown in \Cref{fig:dimension}, most teams opted for 2D models, which is a common and effective strategy for CMR due to the anisotropic resolution of typical SAX acquisitions.

\begin{center}
  \centering
  \includegraphics[width=\linewidth]{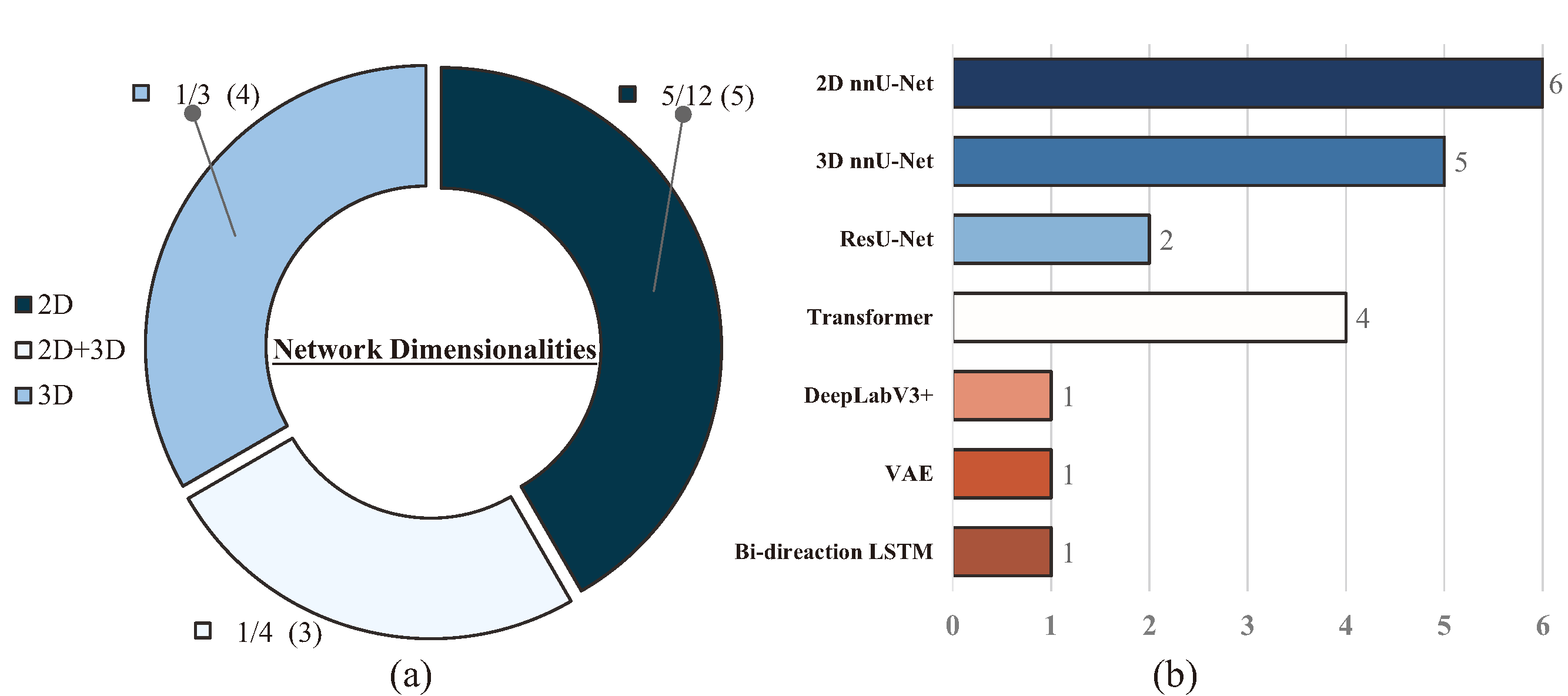}
  \captionof{figure}{Overview of (a) model dimensionalities and (b) network architectures adopted by the participating teams in the RCS task.}
  \label{fig:dimension}
\end{center}

In terms of training paradigms, most teams employed an end-to-end supervised framework, commonly using a combination of Dice and cross-entropy losses. Several teams explored more advanced strategies, such as multi-stage training with different loss functions~\citep{mora-rubio-deep-2022}, two-stage coarse-to-fine segmentation~\citep{li-automatic-2022, qayyum_automatic_2022}, and multi-task learning~\citep{grzeszczyk_multi-task_2022, kou_3d_2022}. Pre-training on large external datasets (both natural images~\citep{5206848} and medical CMR images~\citep{Radau_Lu_Connelly_Paul_Dick_Wright2009,petitjean_right_2015,7029623,ADSB}) and the use of semi-supervised techniques with pseudo-labeling were also feasible strategise to improve robustness.

\subsubsection{Ensemble strategies}
Model ensembling was a widely adopted and highly effective strategy. Common approaches included aggregating predictions from models trained across different cross-validation folds~\citep{gong-robust-2022,grzeszczyk_multi-task_2022,qayyum_automatic_2022} and combining different network dimensionalities (e.g., 2D and 3D models~\citep{yang-deep-2022,ranem_detecting_2022}), or architectures (e.g., CNNs and Transformers~\citep{gong-robust-2022}) to leverage their complementary strengths. Some teams also developed more specialized ensemble frameworks, such as combining models trained with different data sampling strategies to better handle the challenging basal and apical slices~\citep{amirrajab-cardiac-2022}.

\subsubsection{Data preprocessing and postprocessing}
Nearly all participants utilized a comprehensive preprocessing pipeline. This typically involved spatial normalization (resampling to isotropic resolution, resizing, cropping, padding, and orientation normalization) and intensity normalization (z-score, min-max scaling, histogram standardization, and intensity clipping). Several top teams leveraged the automated preprocessing pipeline of the nnU-Net framework~\citep{nnUNet}. Postprocessing was less commonly applied but was utilized by four teams to refine predictions. Such techniques included selecting the largest connected component for each anatomical structure (either slice-wise or volume-wise) or employing the postprocessing pipeline provided by nnU-Net.

\subsubsection{Data augmentation}
Extensive data augmentation was a universal feature of the submitted methods. In addition to standard spatial and intensity-based transformations, many teams implemented augmentations specifically designed to simulate motion artifacts. These included k-space-based artifact simulation and adversarial augmentation techniques like AdvChain~\citep{chen_enhancing_2022}, where the model is trained to be robust against worst-case perturbations. Test-Time Augmentation (TTA) was also employed by some teams to improve the robustness of the final predictions.

\subsection{Details of RCS algorithms}
\subsubsection{S1. UA-SVCC} 
UA-SVCC~\citep{mora-rubio-deep-2022} employed a modified version of the nnU-Net framework~\citep{isensee_nnu-net_2021}, which has previously demonstrated strong performance in CMR segmentation challenges. The authors utilized the official nnU-Net implementation~\citep{nnUNet} to handle preprocessing tasks such as resampling, normalization, patch size configuration, and the adjustment of training hyperparameters (e.g., batch size). These parameters were tailored according to the characteristics of the input data and the available computational resources.

This method adopted a multi-stage training strategy. In the initial stage, only the cross-entropy loss was used. In the second stage, both cross-entropy and Dice losses were combined. In the final stage, only the Dice loss was employed. This progression allows the network to first prioritize accurate class distribution (via cross-entropy loss) and then refine the spatial alignment of segmentation regions (via Dice loss). As a result of this training strategy, the method achieved state-of-the-art performance on Task 2 during the testing phase.

\subsubsection{S2. Med-Air}
Team Med-Air tied for first place in Task 2 due to its comparable performance with S1.UA-SVCC in the final rankings (see \Cref{sec:rank}). This approach proposed a comprehensive ensemble framework that integrated multiple architectures, including 2D nnU-Net, 3D nnU-Net, Swin-UNETR~\citep{hatamizadeh_swin_2022}, and Swin-UNet~\citep{cao_swin-unet_2023}, together with a diverse collection of publicly available datasets used as external resources.

For the transformer-based models, the encoder of Swin-UNETR was pretrained using four external datasets (SCD~\citep{Radau_Lu_Connelly_Paul_Dick_Wright2009}, RVSC~\citep{petitjean_right_2015}, LASB'13~\citep{7029623}, and ADSB~\citep{ADSB}) via multi-task self-supervised learning methods~\citep{Tang_2022_CVPR}. The encoder of Swin-UNet was pretrained on ImageNet~\citep{5206848}. Subsequently, all candidate models were further pretrained on publicly available cardiac segmentation datasets, including ACDC~\citep{ACDC}, M\&Ms-1~\citep{9458279}, M\&Ms-2~\citep{10103611}, MyoPS~\citep{8458220}, and MS-CMRSeg~\citep{zhuang_multivariate_2016}. To enhance the model's robustness to diffeomorphic deformations and spatial transformations, the AdvChain framework~\citep{chen_enhancing_2022} was employed in conjunction with random data augmentation in an iterative training scheme.

These four network architectures, incorporating both pretraining and adversarial augmentations, were trained under a 5-fold cross-validation setting on the training set to produce the average ensemble prediction. Evaluation on the validation dataset demonstrated that pretraining led to notable improvements in the segmentation accuracy of the LV and MYO, while AdvChain contributed substantial performance gains for the RV. As a result, this approach achieved first place and yielded the best DSC scores for the MYO and RV regions.

\subsubsection{S3. CMR.Love.LHND}
CMR.Love.LHND\citep{yang-deep-2022} ranked second in Task 2 and investigated the use of two different network architectures. Initially, they employed a combination of a bidirectional convolutional LSTM (Bi-ConvLSTM)~\citep{bai2020population} and a 2D U-Net, inspired by previous research~\citep{8363618} which demonstrated that recurrent neural networks (RNNs) can capture features across consecutive frames of the cardiac cycle, potentially improving segmentation performance by leveraging spatial continuity along the slice dimension. After comparing the performance of Bi-ConvLSTM and nnU-Net, they ultimately adopted the nnU-Net framework, which integrates both 2D and 3D models in an ensemble configuration. This network was further enhanced by extensive data augmentation strategies, including aggressive Gaussian noise, Gaussian blur, and modifications to brightness, contrast, and gamma correction. This comprehensive setup contributed to the model's robustness and obtained a second-place ranking on the testing set.

\subsubsection{S4. Tewodrosw}
Tewodrosw~\citep{arega-automatic-2022} also tied for second place and employed nnU-Net as the baseline segmentation network. This approach proposed a moderate k-space-based motion artifact augmentation strategy and introduced a hybrid loss function, termed DicePolyCE loss, which combines a region-based Dice loss with a polynomial variant of the cross-entropy loss. This design enhances segmentation performance and improves the model's generalization capability for cardiac MR segmentation under motion artifact conditions.

The augmentation strategy used in this method was categorized into three levels: light augmentation, moderate motion artifact augmentation (moderate MAA), and heavy motion artifact augmentation (heavy MAA). These levels were defined based on the number of simulated motion artifacts generated using the TorchIO library~\citep{perez-garcia_torchio_2021}. Extensive experiments demonstrated that moderate MAA achieved the best performance, whereas light augmentation yielded suboptimal results, and heavy MAA led to excessive image distortion. Additionally, k-space motion artifacts were applied to randomly selected high-quality training images. These synthesized images, combined with supplementary data from the external ACDC dataset~\citep{ACDC}, were incorporated into the training set to enhance sample diversity. This strategy further improved segmentation performance, contributing to the method's second-place ranking in Task 2.

\subsubsection{S5. OpenGTN}
Team OpenGTN~\citep{amirrajab-cardiac-2022} secured third place in Task 2 by developing an ensemble framework based on modified 2D nnU-Net models tailored to address data variability and the segmentation challenges of the basal and apical heart regions, particularly under motion artifacts. To augment the training dataset with motion artifact, respiratory motion is simulated by applying sinusoidal translations to artifact-free images before transforming them into the Fourier domain, allowing adjustment of artifact severity via parameters controlling the period and amplitude of the motion, the method incorporated two external datasets, namely M\&Ms-1~\citep{9458279} and M\&Ms-2~\citep{10103611}.

The final heart cavity segmentation was generated by ensembling predictions from three distinct training strategies: (1) training a 2D nnU-Net on the full dataset with simulated motion artifacts, (2) training another model using non-homogeneous batch sampling, and (3) combining three region-specific models specialized for the basal, mid-ventricular, and apical slices.

To address the segmentation difficulty in basal and apical slices, especially in the presence of motion artifact, the non-homogeneous batch sampling method was employed to oversample these critical slices within each mini-batch. Moreover, Focal-Tversky loss~\citep{8759329} was utilized for region-specific training of the apical and basal areas, focusing on better boundary delineation and class imbalance. These strategies collectively led to improved segmentation performance across all cardiac regions and enabled the team to achieve third place in the final rankings.

\subsubsection{S6. Abdul}
Abdul~\citep{qayyum_automatic_2022} proposed a two-stage segmentation framework to enhance performance through a combination of generalization and pseudo-label supervision. In the first stage, a 3D-ResUNet was employed, incorporating lightweight 3D convolutional blocks in the encoder, residual modules~\citep{He_Res_2016} in the skip connections, and a three-level deep supervision mechanism~\citep{pmlr-v38-lee15a}. This network generated pseudo labels on the validation set using an ensemble of models trained via cross-validation. In the second stage, an nnU-Net was trained using both the pseudo labels and original training data. The 3D-ResUNet architecture featured an encoder-decoder structure with residual blocks, 3D strided convolutions for spatial downsampling, and 3D transposed convolutions for upsampling. Feature maps from the encoder were concatenated with those in the decoder to preserve semantic context, while deep supervision guided the network by computing aggregated losses at multiple levels. The modified nnU-Net was trained with a batch size of $96\times160\times160$ over 500 epochs using one-fold cross-validation.

\subsubsection{S7. Philips\_CTS}
Team Philips\_CTS~\citep{li-automatic-2022} adopted the 2D nnU-Net as the backbone architecture, which was deemed more suitable due to the relatively large slice thickness in the dataset. To improve the segmentation performance, particularly for small anatomical structures, the team proposed a sequential, cascaded two-stage segmentation framework.

In the first stage, a nnU-Net was trained to predict a binary mask encompassing all three structures (i.e., LV, MYO, and RV). Based on this output, new input images were generated by zeroing out the pixels outside the binary mask. These masked images, along with the original three-class labels, were then used to train a second model. The input to the second-stage model was obtained by multiplying the original image by the binary mask from the first stage, and the output consisted of three separate masks corresponding to the target structures.

To mitigate data scarcity and leverage additional information from unlabeled data, the model was further trained using pseudo labels derived from unlabeled samples. This strategy enhanced the model's performance on the validation set, ultimately securing $5_{th}$ place in Task 2.

\subsubsection{S8. MI-IST\_DA}
Team MI-IST\_DA~\citep{ranem_detecting_2022} adopted a 2D ViT U-Net V2 framework~\citep{Ranem_2022_CVPR}, derived from the Lifelong nnU-Net~\citep{gonzalez_lifelong_2023} and based on a Vision Transformer~\citep{dosovitskiy_image_2020} backbone, for cardiac segmentation. This architecture was chosen after a comparative evaluation against nnU-Net models in both 2D and 3D configurations, where ViT U-Net demonstrated the highest scores across all segmentation labels and superior robustness to motion artefacts. Visualization analyses further showed that ViT U-Net generated more complete segmentations of the MYO and LV compared to 2D and 3D nnU-Net models in local evaluations, supporting the hypothesis that the self-attention mechanism of Transformers allows the model to focus more effectively on cardiac structures under motion artefacts. For the final submission, an ensemble of three models -- namely 2D ViT U-Net, 2D nnU-Net, and 3D nnU-Net -- was used to combine their complementary strengths. Enhanced by the dynamic preprocessing and postprocessing pipeline of nnU-Net, this ensemble ultimately ranked $6_{th}$ in Task 2.

\subsubsection{S9. HAHA2}
Team HAHA2~\citep{ma_semi-supervised_2022} developed a semi-supervised segmentation framework called Confidence-Aware Cross Pseudo Supervision (CACPS), based on a semi-supervised domain generalization strategy~\citep{2022arXiv220108657Y}, to improve the quality of pseudo labels. This approach treats different respiratory motions as distinct image domains and employs Fourier transformation for data augmentation. Specifically, paired images from the source domain are randomly sampled and transformed into the frequency domain via Fourier transformation. A weighted combination of their amplitude spectra, together with phase information from one of the images, is then computed to synthesize new training images. These generated and the original images are fed into a dual-path confidence-aware cross pseudo supervision network inspired by~\citep{2021arXiv210601226C}, with DeepLabV3+~\citep{2017arXiv170605587C} as the backbone. The proposed CACPS framework outperformed the baseline DeepLabV3+, and further improvements were observed by ensembling two CACPS networks with different backbones (i.e., DeepLabV3+ and ResNet-101).

\subsubsection{S10. ML-Labs}
Team ML-Labs~\citep{garcia-cabrera_cardiac_2022} implemented a ResNet-based U-Net~\citep{ronneberger_u-net_2015} architecture that employs ResNet-101~\citep{He_Res_2016} as the encoder, with weights pre-trained on the ImageNet classification task. The approach integrates four effective data augmentation techniques to enhance robustness. These techniques include random motion simulation following the method in \citep{pmlr-v102-shaw19a}, random ghosting to simulate the effect along the phase-encoded direction, and altering the intensity of imaged structures. It also employs random bias field transformation to introduce low-frequency intensity inhomogeneity, modeled as a linear combination of polynomial functions (as described in \citep{811270}), and random gamma transformation to adjust image contrast by raising pixel values to a random power. Extended experiments demonstrated that tripling the intensity of random motion augmentation, compared to the other techniques, results in model that is more resilient to respiratory motion artifacts, thereby improving segmentation performance.

\subsubsection{S11. Sano}
Team Sano~\citep{grzeszczyk_multi-task_2022} proposed a multi-task learning framework that concurrently performs motion artifact classification and CMR segmentation in a single forward pass. Their approach is based on a modified multi-task architecture inspired by Swin UNETR (Swin U-Net Transformers), following the design introduced by \citep{hatamizadeh_swin_2022}. In this framework, segmentation is treated as the primary task, while IQA serves as an auxiliary task. The final prediction is obtained through an ensemble of five networks trained using 5-fold stratified cross-validation. Stratification is performed based on the classification labels to ensure an even distribution of motion artifacts across all folds. This method achieved competitive results in terms of DSC and HD95.

\begin{figure*}[htbp]
  \centering
  \includegraphics[width=\linewidth]{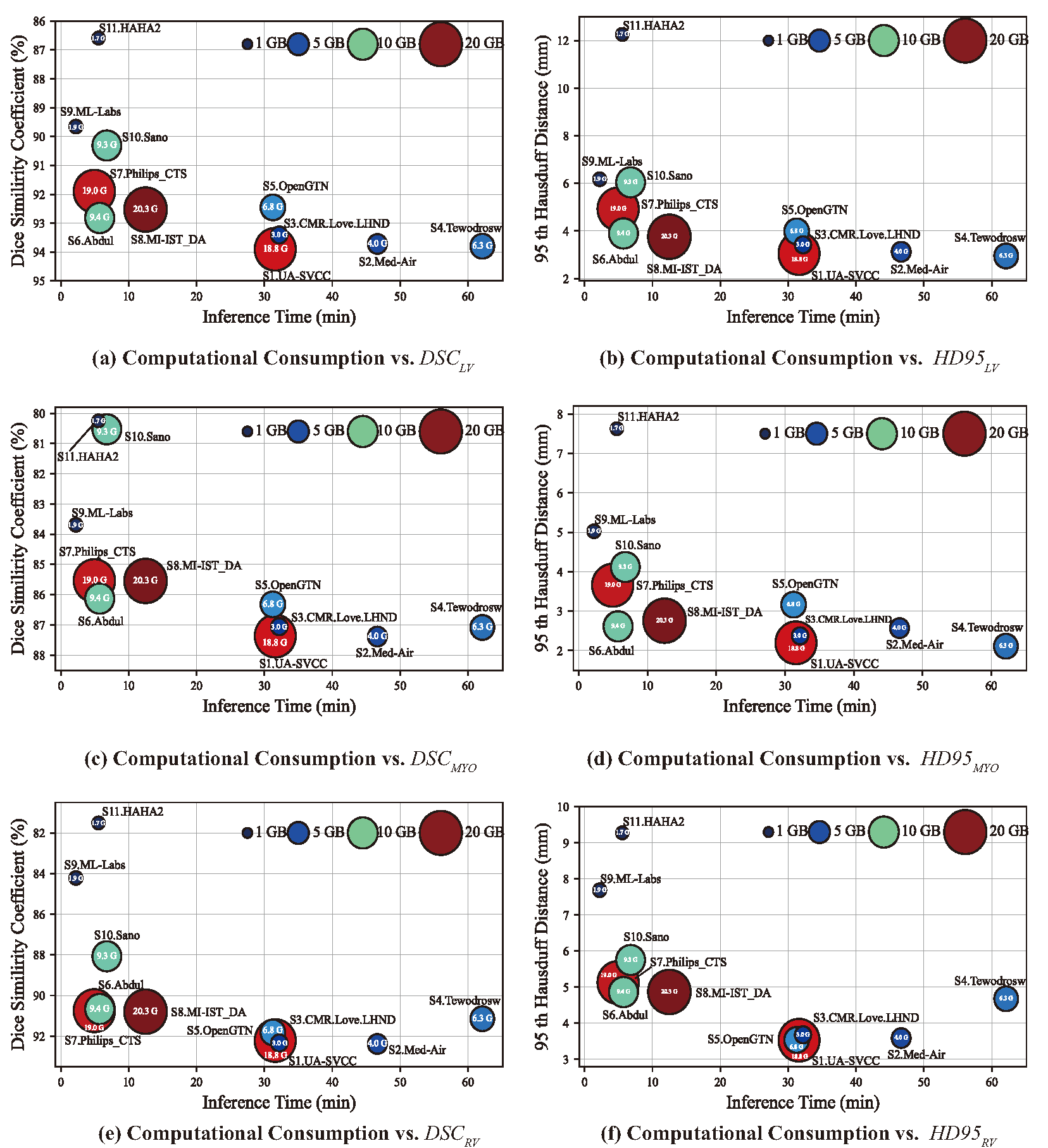}
  \caption{Analysis of computational efficiency versus segmentation performance for the top 10 teams in the RCS task. The x-axis of each plot represents the total inference time, while the y-axis represents the segmentation metric (DSC or HD95) for the LV, MYO, and RV. The size of each data point corresponds to the peak GPU VRAM usage.}
  \label{fig:efficiency}
\end{figure*}

\subsubsection{S12. sots}
Team SOTS~\citep{kou_3d_2022} also proposed a multi-task learning framework that integrates DenseBiasNet~\citep{HE2020101722} with a Variational Auto-Encoder (VAE) inspired by~\citep{10.1007/978-3-030-11726-9_28}. In this framework, DenseBiasNet serves as the primary branch for segmentation, while the VAE functions as an auxiliary branch to enhance representation learning.

A shared encoder with dense bias connections from DenseBiasNet is employed to fuse multi-scale and multi-receptive field features, providing rich contextual representations for both segmentation and reconstruction tasks. The VAE takes as input a low-resolution image derived from the encoder output of DenseBiasNet and maps it to a low-dimensional latent space to reconstruct the original image. The segmentation task is supervised using a combination of Soft Dice loss~\citep{7785132} and cross-entropy loss, whereas the reconstruction task is guided by an L2 loss.

Collaborative training of the VAE and DenseBiasNet branches yields improved segmentation performance compared to using DenseBiasNet alone. Notably, local validation results indicate that as the severity of respiratory motion artifacts increases, the segmentation quality of both the LV and MYO becomes more dominant, suggesting enhanced robustness under artifact-heavy conditions.

\subsection{Computational efficiency analysis}
\label{sec:efficiency}
Beyond segmentation accuracy, the computational efficiency of the algorithms is a critical factor for their potential clinical translation. We evaluated the inference time and peak GPU memory usage of all submitted Docker containers on a standardized hardware platform. The results are summarized in \Cref{tab:consumption_task1} and \Cref{tab:consumption_task2}. 

\begin{figure*}[t]
  \vspace{-0.4cm}
  \centering
  \includegraphics[width=\linewidth]{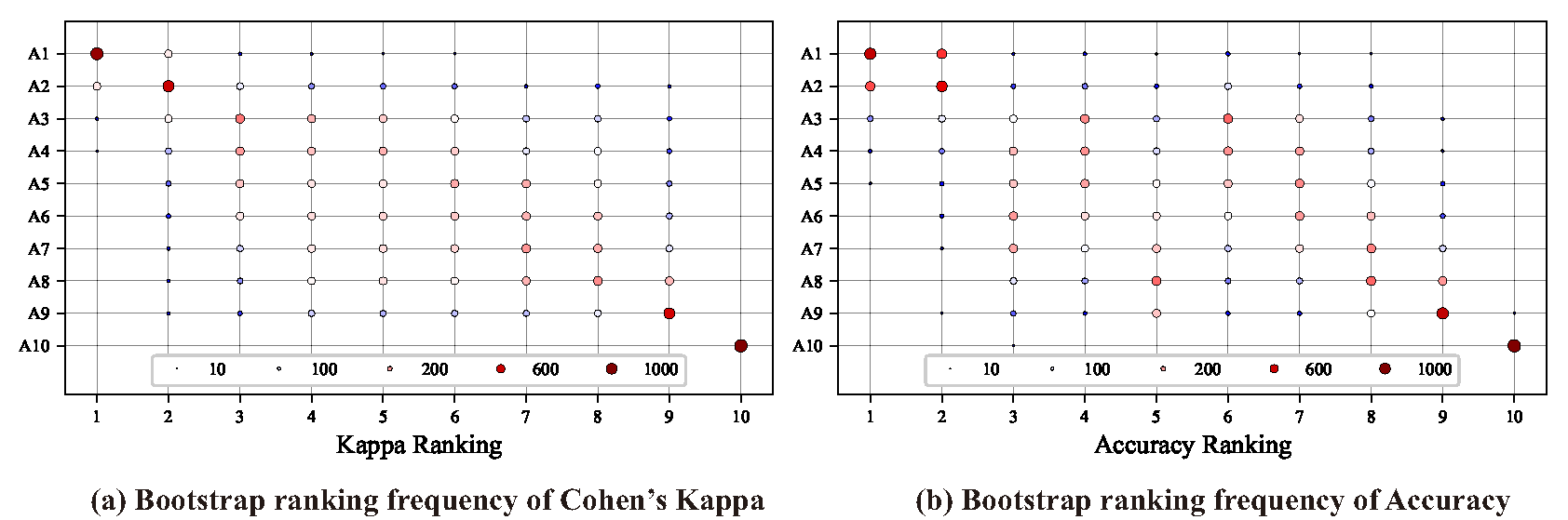}
  \caption{Matrix bubble chart of ranking statistics for IQA task (Task 1) based on 1,000 bootstrap samples. (a) the bootstrap ranking frequencies for Cohen's Kappa, while (b) the bootstrap ranking frequencies for Accuracy. In this visualization, the area of each data node indicates its ranking frequency across the 1,000 bootstrap samples. The corresponding color is assigned using a value-based colormap that resembles the seismic colormap.}
  \label{fig:IQA_rank}
  \vspace{-0.4cm}
\end{figure*}

\begin{center}
  \centering
  \captionof{table}{The computation consumption for the IQA task on the testing set. The \textbf{Max VRAM} represents the allocated peak GPU VRAM, and the \textbf{Inference time} is the total processing time, which includes data loading, preprocessing, model inference, post-processing, and data saving. These measurements were taken using 120 testing images during the testing phase on our testing platform. None represents the algorithm is specified running in cpu mode.}
  \label{tab:consumption_task1}%
  \resizebox{\linewidth}{!}{
  \begin{tabular}{lcrr}
    \toprule
    \textbf{TID} & \multicolumn{1}{c}{\textbf{Team Name}} & \multicolumn{1}{c}{\textbf{Max VRAM}} & \multicolumn{1}{c}{\textbf{Inference Time}} \\
    \midrule
    \textbf{A1} & UON\_IMA & 2.46 GB & 11.94 min \\
    \textbf{A2} & Philips\_CTS & 19.08 GB & 14.98 min \\
    \textbf{A3} & OpenGTN & -     & 10.12 min \\
    \textbf{A4} & issun & 4.41 GB & 78.44 min \\
    \textbf{A5} & Tewodrosw & 1.77 GB & 1.33 min \\
    \textbf{A6} & CMR.Love.LHND & 1.58 GB & 1.78 min \\
    \textbf{A7} & MI-IST\_DA & 20.33 GB & 12.36 min \\
    \textbf{A8} & UA-SVCC & 2.49 GB & 0.45 min \\
    \textbf{A9} & Abdul & 2.99 GB & 1.17 min \\
    \textbf{A10} & Sano  & 9.31 GB & 6.06 min \\
    \bottomrule
  \end{tabular}%
  }
\end{center}%

For the RCS task, the total inference time for the 120 test cases ranged from 1.19 to 62.12 minutes, with peak GPU VRAM consumption ranging from 1.69 GB to 20.27 GB. As illustrated in \Cref{fig:efficiency}, there was a clear trade-off between performance and efficiency. The top-ranked algorithms from UA-SVCC and Med-Air, while achieving the highest accuracy, were also among the more computationally intensive solutions. Conversely, algorithms that prioritized faster inference often did so at the cost of segmentation accuracy. These results highlight the ongoing challenge of developing models that are both highly accurate and computationally efficient enough for seamless integration into clinical workflows.

\begin{center}
  \centering
  \captionof{table}{The computation consumption for the segmentation task on the testing set. The \textbf{Max VRAM} represents the peak allocated GPU VRAM, and the \textbf{Inference time} is the total processing time, which includes data loading, preprocessing, model inference, post-processing, and data saving. These measurements were taken using 120 testing images during the testing phase on our testing platform.}
  \label{tab:consumption_task2}%
  \resizebox{\linewidth}{!}
  {
      \begin{tabular}{lcrr}
      \toprule
      \textbf{TID} & \multicolumn{1}{c}{\textbf{Team Name}} & \multicolumn{1}{c}{\textbf{Max VRAM}} & \multicolumn{1}{c}{\textbf{Inference Time}} \\
      \midrule
      \textbf{S1} & UA-SVCC & 18.81 GB & 31.60 min \\
      \textbf{S2} & Med-Air & 4.01 GB & 46.65 min \\
      \textbf{S3} & CMR.Love.LHND   & 2.98 GB & 32.19 min \\
      \textbf{S4} & Tewodrosw & 6.32 GB & 62.12 min \\
      \textbf{S5} & OpenGTN & 6.80 GB & 31.25 min \\
      \textbf{S6} & Abdul & 9.41 GB & 5.73 min \\
      \textbf{S7} & Philips\_CTS & 19.02 GB & 4.94 min \\
      \textbf{S8} & MI-IST\_DA & 20.27 GB & 12.49 min \\
      \textbf{S9} & ML-Labs & 1.90 GB & 2.22 min \\
      \textbf{S10} & Sano  & 9.31 GB & 6.78 min \\
      \textbf{S11} & HAHA2  & 1.69 GB & 5.55 min \\
      \textbf{S12} & sots  & 6.06 GB & 1.19 min \\
      \bottomrule
      \end{tabular}%
  }
\end{center}%

\subsection{Ranking analysis}
\subsubsection{Ranking of IQA task}

The final ranking for the IQA task was determined by the linearly weighted Cohen's Kappa coefficient, a metric chosen for its robustness to class imbalance compared to standard accuracy. As reported in \Cref{tab:task1_results}, submissions were sorted in descending order based on their Kappa score on the held-out test set. To evaluate the stability of this ranking, we performed a bootstrap analysis (1,000 samples) for both the Kappa and accuracy metrics. The results, visualized in \Cref{fig:IQA_rank}, demonstrate that Kappa provides a more stable and reliable ranking. The ranking frequency distribution for Kappa is highly concentrated along the main diagonal, indicating that the top-performing teams consistently maintained their high ranks across the bootstrap samples. In contrast, the ranking distribution based on accuracy was significantly more dispersed, with greater variance and a higher degree of randomness, reinforcing the choice of Kappa as the primary evaluation metric.

\subsubsection{Ranking of RCS task}
\label{sec:rank}
For the RCS task, we employed a robust \textit{rank-then-aggregate} strategy to determine the final standings. This approach ensures that the final ranking reflects consistent performance across all evaluation criteria, rather than being skewed by exceptional performance on a single metric or structure. For each of the 120 test cases, every participating team was ranked based on six individual metrics: the Dice Similarity Coefficient (DSC) and the 95\% Hausdorff Distance (HD95) for each of the three anatomical structures (LV, MYO, and RV). The final ranking score for each team was then calculated as the mean of these individual ranks across all test cases. To determine if the differences between the final ranking scores were statistically significant, we performed pairwise comparisons using the two-sided Wilcoxon signed-rank test, with p-values corrected for multiple comparisons using the Holm-Bonferroni procedure. As shown in the significance matrix in \Cref{fig:Seg_Rank}, this analysis revealed no significant difference between the top two teams (S1.UA-SVCC and S2.Med-Air), resulting in a tie for first place. Similarly, teams S3.CMR.Love.LHND and S4.Tewodrosw were found to have statistically comparable performance, resulting in a tie for second place.

\begin{center}
  \centering
  \includegraphics[width=\linewidth]{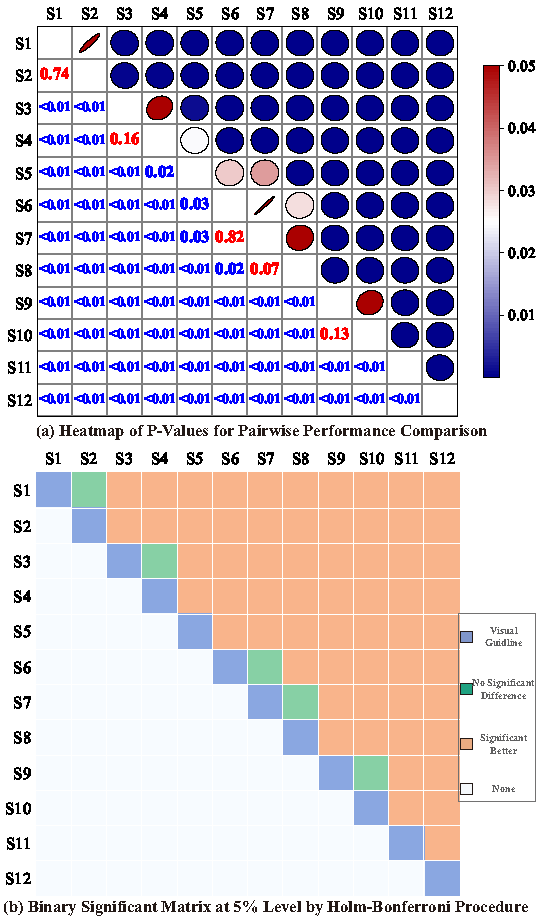}
  \captionof{figure}{Statistical analysis of the RCS ranking. (a) p-value heatmap from the pairwise Wilcoxon signed-rank tests. The shrinkage of each ellipse corresponds to the p-value. (b) binary significance matrix at the 5\% level after Holm-Bonferroni correction. Orange indicates the row significantly outperformed the column, while cyan indicates no significant difference.}
  \label{fig:Seg_Rank}
\end{center}

\section{Discussion}
\begin{table*}[htbp]
  \centering
  \caption{Subgroup analysis of segmentation performance for top 3 methods under different IQA scores. The reported data are in the format of mean $\pm$ standard deviation. $\uparrow$ represents higher is better,  $\downarrow$ represents lower is better. Bold represent the best performance of corresponding IQA score.}
  \resizebox{\linewidth}{!}{
    \begin{tabular}{clcccccccc}
    \toprule
    \multirow{2}[4]{*}{IQA Score} & \multicolumn{1}{c}{\multirow{2}[4]{*}{Methods}} & \multicolumn{3}{c}{DSC (\%)$\uparrow$} &       &       & \multicolumn{3}{c}{HD95 ($mm$)$\downarrow$} \\
\cmidrule{3-5}\cmidrule{7-10}          &       & LV    & MYO   & RV    &       &       & LV    & MYO   & RV \\
    \midrule
    \multirow{5}[2]{*}{1} & S1. UA-SVCC & \textbf{94.63 $\pm$ 3.27} & 87.82 $\pm$ 4.17 & 93.14 $\pm$ 5.59 &       &       & \textbf{2.52 $\pm$ 3.19} & 2.27 $\pm$ 2.50 & 2.95 $\pm$ 3.21 \\
          & S2. Med-Air & 94.49 $\pm$ 4.04 & \textbf{88.10 $\pm$ 3.71} & 93.20 $\pm$ 5.53 &       &       & 2.65 $\pm$ 3.57 & 2.51 $\pm$ 2.92 & 3.07 $\pm$ 3.95 \\
          & S3. CMR.Love.LHND & 94.22 $\pm$ 3.47 & 87.57 $\pm$ 4.26 & \textbf{93.45 $\pm$ 3.82} &       &       & 2.96 $\pm$ 3.58 & 2.31 $\pm$ 2.49 & \textbf{2.82 $\pm$ 2.83} \\
          & S4. Tewodrosw & 94.35 $\pm$ 3.53 & 87.73 $\pm$ 3.84 & 92.16 $\pm$ 5.48 &       &       & 2.75 $\pm$ 3.41 & \textbf{2.02 $\pm$ 1.85} & 4.22 $\pm$ 3.84 \\
          & S5. OpenGTN & 93.10 $\pm$ 5.18 & 86.95 $\pm$ 3.69 & 92.44 $\pm$ 5.64 &       &       & 3.73 $\pm$ 4.53 & 3.09 $\pm$ 3.23 & 3.17 $\pm$ 3.04 \\
    \midrule
    \multirow{5}[2]{*}{2} & S1. UA-SVCC & 93.21 $\pm$ 3.95 & \textbf{86.94 $\pm$ 2.40} & 91.33 $\pm$ 6.36 &       &       & 3.55 $\pm$ 3.77 & \textbf{2.15 $\pm$ 1.69} & 4.08 $\pm$ 3.76 \\
          & S2. Med-Air & 92.98 $\pm$ 4.44 & 86.72 $\pm$ 2.65 & \textbf{91.59 $\pm$ 5.43} &       &       & 3.58 $\pm$ 3.80 & 2.65 $\pm$ 2.50 & 4.10 $\pm$ 3.71 \\
          & S3. CMR.Love.LHND & 92.63 $\pm$ 4.45 & 86.56 $\pm$ 2.75 & 91.28 $\pm$ 4.60 &       &       & 3.87 $\pm$ 3.84 & 2.45 $\pm$ 2.06 & 4.49 $\pm$ 3.67 \\
          & S4. Tewodrosw & \textbf{93.26 $\pm$ 3.91} & 86.49 $\pm$ 2.40 & 90.16 $\pm$ 6.17 &       &       & \textbf{3.17 $\pm$ 3.38} & 2.21 $\pm$ 1.72 & 5.15 $\pm$ 4.14 \\
          & S5. OpenGTN & 91.81 $\pm$ 6.01 & 85.70 $\pm$ 3.22 & 91.19 $\pm$ 6.06 &       &       & 4.27 $\pm$ 4.36 & 3.27 $\pm$ 3.14 & \textbf{3.91 $\pm$ 3.59} \\
    \midrule
    \multirow{5}[2]{*}{3} & S1. UA-SVCC & \textbf{91.76 $\pm$ 5.05} & \textbf{83.12 $\pm$ 6.49} & 87.25 $\pm$ 6.21 &       &       & 3.95 $\pm$ 4.21 & 4.62 $\pm$ 4.84 & 7.45 $\pm$ 5.69 \\
          & S2. Med-Air & 91.01 $\pm$ 5.72 & 82.85 $\pm$ 6.09 & 87.97 $\pm$ 6.66 &       &       & 4.86 $\pm$ 5.55 & 5.06 $\pm$ 5.40 & 6.84 $\pm$ 5.77 \\
          & S3. CMR.Love.LHND & 91.23 $\pm$ 5.01 & 82.07 $\pm$ 6.59 & 87.63 $\pm$ 5.77 &       &       & 4.56 $\pm$ 4.28 & 5.00 $\pm$ 4.77 & 6.39 $\pm$ 4.38 \\
          & S4. Tewodrosw & 91.57 $\pm$ 4.78 & 83.09 $\pm$ 6.26 & 88.49 $\pm$ 5.55 &       &       & \textbf{3.94 $\pm$ 3.54} & \textbf{3.54 $\pm$ 2.73} & 5.75 $\pm$ 4.83 \\
          & S5. OpenGTN & 90.68 $\pm$ 5.51 & 81.74 $\pm$ 5.72 & \textbf{89.03 $\pm$ 5.13} &       &       & 4.35 $\pm$ 3.51 & 4.45 $\pm$ 3.24 & \textbf{5.42 $\pm$ 3.59} \\
    \bottomrule
    \end{tabular}%
  }
  \label{tab:segment_subgroup}%
\end{table*}%

This challenge was designed to establish the first benchmark for assessing the impact of respiratory motion on CMR image quality and the robustness of automated segmentation models. In this section, we analyze the performance of the submitted algorithms, contextualize the results, and discusse the implications for future research. In particular, we further investigate the downstream effects of motion artifacts on clinical imaging biomarkers, and compare the top-performing algorithms against human expert performance.

\subsection{Performance on the IQA task}
The challenge results demonstrate that automated IQA is a feasible but challenging task. The top-performing algorithm achieved a Cohen's Kappa score of 0.631, indicating a good level of agreement with the expert radiologists' assessment. This level of performance suggests that automated tools could be reliably used to flag images with motion artifacts in a clinical workflow.

However, the overall performance landscape reveals significant room for improvement. While the winning method showed a high level of consistency, the majority of submitted algorithms achieved only moderate agreement (Kappa scores between 0.41 and 0.60), and no method reached an excellent level of agreement ($\kappa > 0.80$). A key finding, illustrated by the confusion matrixes in \Cref{fig:Confusion_mtx}, is that while most algorithms could effectively identify high-quality images (mild motion), they struggled to accurately classify the severity of motion in more degraded images. This indicates that while current models are promising for binary quality control (i.e., usable vs. unusable), fine-grained assessment of artifact severity remains an open research problem.

\subsection{Performance on the RCS task}

For the RCS task, the state-of-the-art deep learning models demonstrated high accuracy on the test set, with the best average DSC scores reaching 93.90\% for the LV, 87.38\% for the MYO, and 92.39\% for the RV. Consistent with previous segmentation challenges, the MYO proved to be the most difficult structure to segment due to its thin anatomy, resulting in lower DSC scores despite often achieving good surface distance metrics (HD95).

\subsubsection{The impact of motion artifacts on segmentation and clinical imaging biomarkers}
A central goal of this challenge was to quantify the impact of motion artifacts on segmentation performance. Our subgroup analysis, presented in \Cref{tab:segment_subgroup}, confirms a clear trend: as image quality degrades (i.e., as the IQA score increases), the segmentation accuracy of all top-performing algorithms decreases. For example, the DSC for LV segmentation for the top-ranked algorithm dropped by over 3 percentage points between the best- and worst-quality images.

\begin{figure*}[htp]
  \vspace{-0.4 cm}
  \centering
  \includegraphics[width=\linewidth]{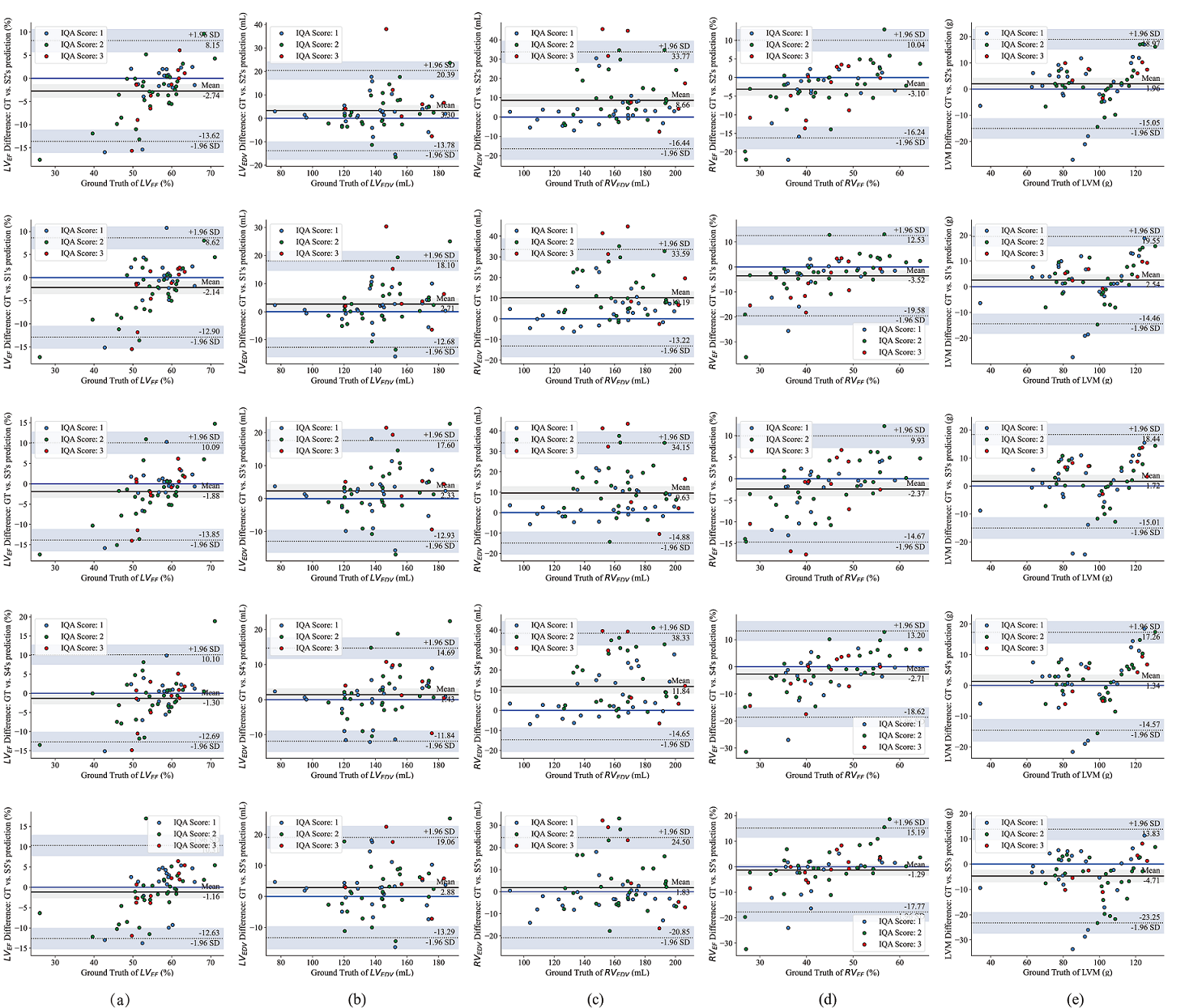}
  \vspace{-0.8 cm}
  \caption{Bland-Altman plots illustrating the agreement between AI-derived and ground-truth clinical biomarkers for the top five RCS algorithms. Each row corresponds to an algorithm, and each column represents a different clinical biomarker: (a) $LV_{EF}$, (b) $RV_{EF}$, (c) $LV_{EDV}$, (d) $LVM$, and (e) $RV_{EDV}$. The solid black line indicates the mean difference, and the dashed lines represent the 95\% limits of agreement.}
  \vspace{-0.4 cm}
  \label{fig:BA_plot}
\end{figure*}

\begin{figure*}[htbp]
  \vspace{-0.2 cm}
  \centering
  \includegraphics[width=\linewidth]{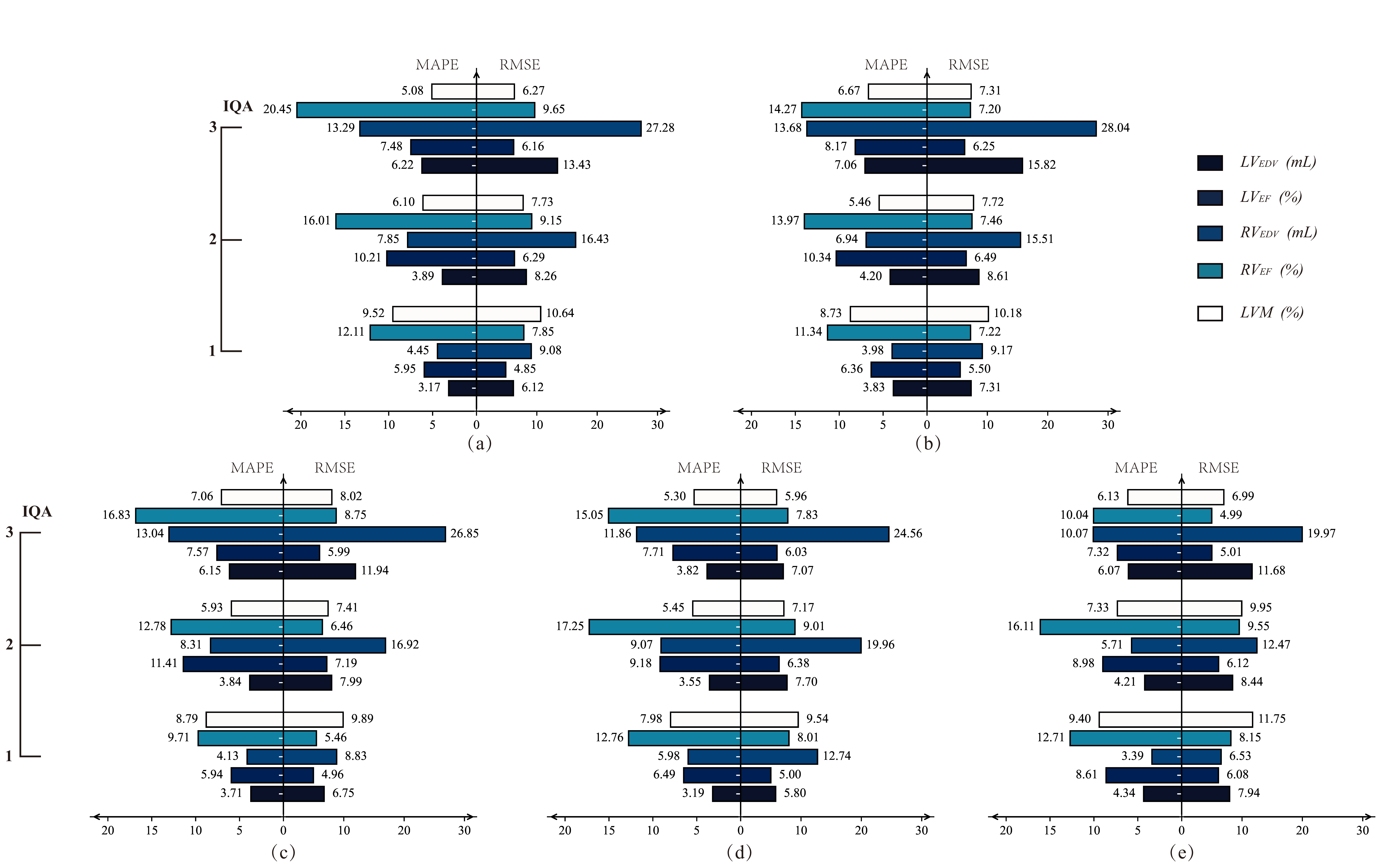}
  \vspace{-0.8 cm}
  \caption{Bidirectional bar charts showing the error in clinical biomarkers for the top five RCS algorithms, stratified by IQA score. The left bars represent the Mean Absolute Percentage Error (MAPE), and the right bars represent the Root Mean Square Error (RMSE).}
  \vspace{-0.4 cm}
  \label{fig:Bi_chart}
\end{figure*}

This degradation in segmentation accuracy has a direct, clinically significant downstream effect. As shown in the Bland-Altman plots in \Cref{fig:BA_plot} and the detailed error analysis in \Cref{tab:cl_metric1,tab:cl_metric2,tab:cl_metric3,tab:cl_metric4,tab:cl_metric5}, the error in derived clinical biomarkers, such as ejection fraction and ventricular volumes, increases substantially as image quality worsens. For instance, the Mean Absolute Error (MAE) for the LV end-diastolic volume ($LV_{EDV}$) more than doubled for the top algorithm when comparing images with mild versus severe motion artifacts as shown in \Cref{fig:Bi_chart}. This finding underscores the critical importance of developing motion-robust segmentation models, as even subtle segmentation inaccuracies in degraded images can lead to significant errors in the clinical parameters used for diagnosis and patient management.

\begin{table*}[htbp]
  \centering
  \begin{minipage}[t]{0.36\textwidth}
  \centering
  \caption{Differences between clinical indicators predicted by S1 and the gold standard across different IQA score levels.}
  \resizebox{5.9 cm}{!}{
    \begin{tabular}{lrrrrrr}
    \toprule
    \multicolumn{1}{c}{\multirow{2}[4]{*}{Metrics}} & \multicolumn{2}{c}{IQA 1} & \multicolumn{2}{c}{IQA 2} & \multicolumn{2}{c}{IQA 3} \\
\cmidrule{2-7}          & \multicolumn{1}{c}{MAE} & \multicolumn{1}{c}{RMSE} & \multicolumn{1}{c}{MAE} & \multicolumn{1}{c}{RMSE} & \multicolumn{1}{c}{MAE} & \multicolumn{1}{c}{RMSE} \\
    \midrule
    $LV_{EDV}$ ($mL$) & 4.51  & 6.12  & 5.84  & 8.26  & 9.55  & 13.43 \\
    $LV_{EF}$ (\%) & 3.14  & 4.85  & 4.77  & 6.29  & 3.93  & 6.16 \\
    $RV_{EDV}$ ($mL$) & 6.69  & 9.08  & 12.68 & 16.43 & 22.10 & 27.28 \\
    $RV_{EF}$ (\%) & 4.67  & 7.85  & 5.91  & 9.15  & 7.79  & 9.65 \\
    $LVM$ ($g$) & 8.23  & 10.70 & 6.53  & 8.11  & 6.24  & 7.88 \\
    \bottomrule
    \end{tabular}%
    \label{tab:cl_metric1}%
  }
  \end{minipage}%
  \hspace{0.03\textwidth}%
  \begin{minipage}[t]{0.36\textwidth}
  \centering
  \caption{Differences between clinical indicators predicted by S2 and the gold standard across different IQA score levels.}
  
  \resizebox{5.9 cm}{!}{
  \begin{tabular}{lrrrrrr}
  \toprule
  \multicolumn{1}{c}{\multirow{2}[4]{*}{Metrics}} & \multicolumn{2}{c}{IQA 1} & \multicolumn{2}{c}{IQA 2} & \multicolumn{2}{c}{IQA 3} \\
\cmidrule{2-7}          & \multicolumn{1}{c}{MAE} & \multicolumn{1}{c}{RMSE} & \multicolumn{1}{c}{MAE} & \multicolumn{1}{c}{RMSE} & \multicolumn{1}{c}{MAE} & \multicolumn{1}{c}{RMSE} \\
  \midrule
  $LV_{EDV}$ ($mL$) & 5.40  & 7.31  & 6.30  & 8.61  & 10.78 & 15.82 \\
  $LV_{EF}$ (\%) & 3.30  & 5.50  & 4.85  & 6.49  & 4.36  & 6.25 \\
  $RV_{EDV}$ ($mL$) & 5.93  & 9.17  & 11.47 & 15.51 & 22.67 & 28.04 \\
  $RV_{EF}$ (\%) & 4.39  & 7.22  & 5.32  & 7.46  & 5.53  & 7.20 \\
  $LVM$ ($g$) & 7.57  & 10.28 & 6.10  & 8.06  & 6.68  & 8.08 \\
  \bottomrule
  \end{tabular}%
  \label{tab:cl_metric2}%
  }
  \end{minipage}%
  
\end{table*}

\begin{table*}[htbp]
  \centering
  \begin{minipage}[t]{0.3\textwidth}
  \centering
  \caption{Differences between clinical indicators predicted by S3 and the gold standard across different IQA score levels.}
  
  \resizebox{5.9 cm}{!}{
  \begin{tabular}{lrrrrrr}
  \toprule
  \multicolumn{1}{c}{\multirow{2}[4]{*}{Metrics}} & \multicolumn{2}{c}{IQA 1} & \multicolumn{2}{c}{IQA 2} & \multicolumn{2}{c}{IQA 3} \\
\cmidrule{2-7}          & \multicolumn{1}{c}{MAE} & \multicolumn{1}{c}{RMSE} & \multicolumn{1}{c}{MAE} & \multicolumn{1}{c}{RMSE} & \multicolumn{1}{c}{MAE} & \multicolumn{1}{c}{RMSE} \\
  \midrule
  $LV_{EDV}$ ($mL$) & 5.19  & 6.75  & 5.78  & 7.99  & 9.40  & 11.94 \\
  $LV_{EF}$ (\%) & 3.11  & 4.96  & 5.47  & 7.19  & 4.07  & 5.99 \\
  $RV_{EDV}$ ($mL$) & 6.32  & 8.83  & 13.63 & 16.92 & 21.60 & 26.85 \\
  $RV_{EF}$ (\%) & 3.78  & 5.46  & 5.04  & 6.46  & 6.59  & 8.75 \\
  $LVM$ ($g$) & 7.70  & 10.27 & 5.70  & 7.32  & 8.19  & 8.87 \\
  \bottomrule
  \end{tabular}%
  \label{tab:cl_metric3}%
  }
  \end{minipage}%
  \hspace{0.03\textwidth}%
  \begin{minipage}[t]{0.3\textwidth}
  \centering
  \caption{Differences between clinical indicators predicted by S4 and the gold standard across different IQA score levels.}
  \label{tab:cl_metric4}%
  \resizebox{5.9 cm}{!}{
  \begin{tabular}{lrrrrrr}
  \toprule
  \multicolumn{1}{c}{\multirow{2}[4]{*}{Metrics}} & \multicolumn{2}{c}{IQA 1} & \multicolumn{2}{c}{IQA 2} & \multicolumn{2}{c}{IQA 3} \\
\cmidrule{2-7}          & \multicolumn{1}{c}{MAE} & \multicolumn{1}{c}{RMSE} & \multicolumn{1}{c}{MAE} & \multicolumn{1}{c}{RMSE} & \multicolumn{1}{c}{MAE} & \multicolumn{1}{c}{RMSE} \\
  \midrule
  $LV_{EDV}$ ($mL$) & 4.44  & 5.80  & 5.29  & 7.70  & 5.88  & 7.07 \\
  $LV_{EF}$ (\%) & 3.43  & 5.00  & 4.54  & 6.38  & 4.10  & 6.03 \\
  $RV_{EDV}$ ($mL$) & 9.53  & 12.74 & 14.92 & 19.96 & 19.52 & 24.56 \\
  $RV_{EF}$ (\%) & 5.14  & 8.01  & 6.56  & 9.01  & 5.67  & 7.83 \\
  $LVM$ ($g$) & 6.94  & 9.47  & 5.92  & 7.50  & 6.09  & 7.44 \\
  \bottomrule
  \end{tabular}%
  }
  \end{minipage}%
  \hspace{0.03\textwidth}%
  \begin{minipage}[t]{0.3\textwidth}
  \centering
  \caption{Differences between clinical indicators predicted by S5 and the gold standard across different IQA score levels.}
  \label{tab:cl_metric5}%
  \resizebox{5.9 cm}{!}{
  \begin{tabular}{lrrrrrr}
  \toprule
  \multicolumn{1}{c}{\multirow{2}[4]{*}{Metrics}} & \multicolumn{2}{c}{IQA 1} & \multicolumn{2}{c}{IQA 2} & \multicolumn{2}{c}{IQA 3} \\
\cmidrule{2-7}          & \multicolumn{1}{c}{MAE} & \multicolumn{1}{c}{RMSE} & \multicolumn{1}{c}{MAE} & \multicolumn{1}{c}{RMSE} & \multicolumn{1}{c}{MAE} & \multicolumn{1}{c}{RMSE} \\
  \midrule
  $LV_{EDV}$ ($mL$) & 6.02  & 7.94  & 6.12  & 8.44  & 9.37  & 11.68 \\
  $LV_{EF}$ (\%) & 4.72  & 6.08  & 4.41  & 6.12  & 4.06  & 5.01 \\
  $RV_{EDV}$ ($mL$) & 4.86  & 6.53  & 9.20  & 12.47 & 16.69 & 19.97 \\
  $RV_{EF}$ (\%) & 4.95  & 8.15  & 6.25  & 9.55  & 4.03  & 4.99 \\
  $LVM$ ($g$) & 8.12  & 12.07 & 6.53  & 8.84  & 8.97  & 11.68 \\
  \bottomrule
  \end{tabular}%
  }
  \end{minipage}
\end{table*}

\subsubsection{Comparison with human expert performance}
\begin{figure*}[htbp]
  \centering
  \includegraphics[width=\linewidth]{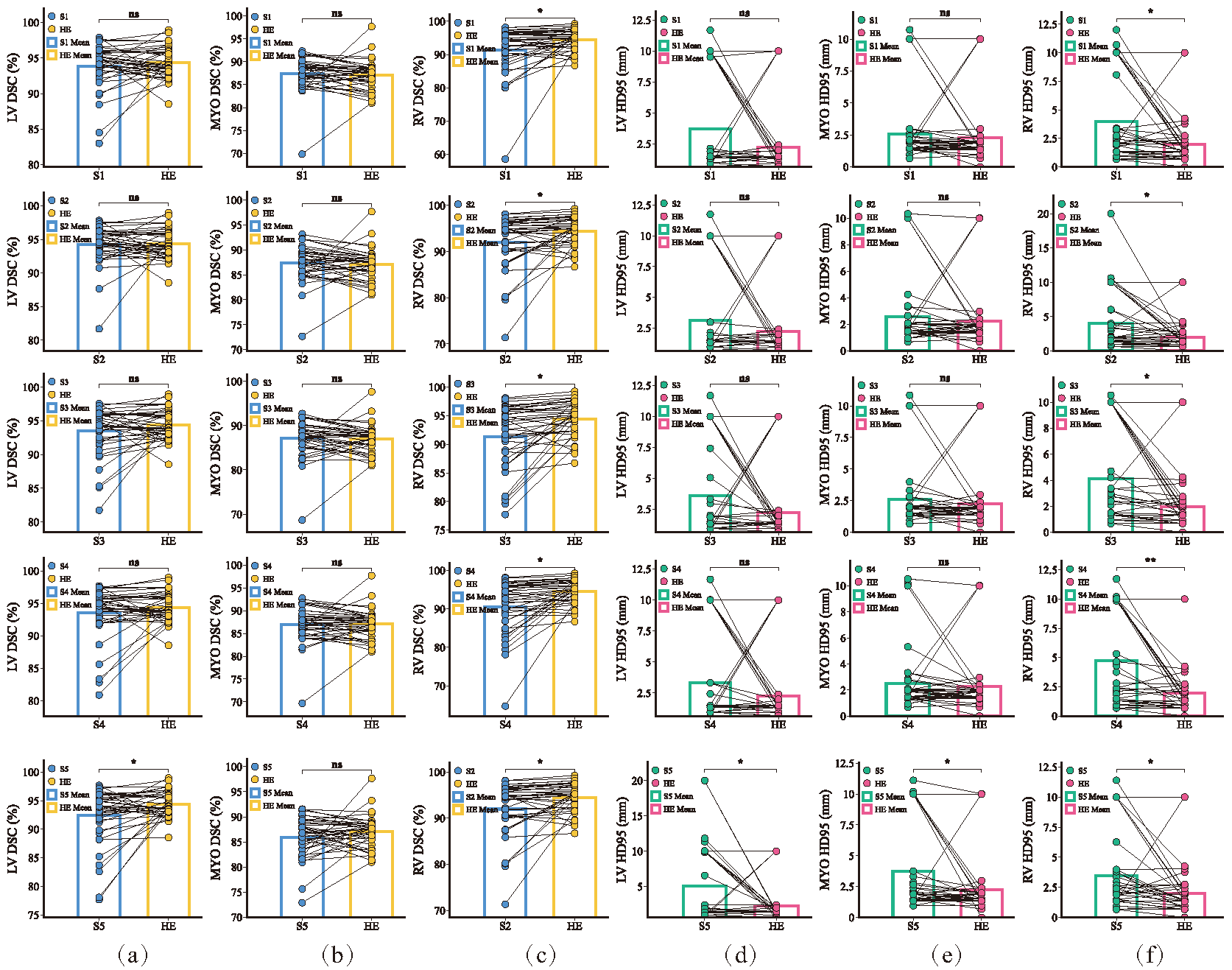}
  \caption{Paired t-test comparison between the top 3 AI models (S1-S5) and a human expert (HE) on 40 CMR volumes. The plots show performance for (a-c) DSC and (d-f) HD95 for the LV, MYO, and RV. \textbf{ns} indicates no significant difference, * indicates p $\le$ 0.05, and ** indicates p $\le$ 0.001.}
  \label{fig:pair-t-test}
\end{figure*}

To contextualize the performance of the automated methods, we conducted an inter-observer validation study. The error analysis of five derived clinical biomarkers calculated from the annotations of human exprts are shown in \Cref{fig:Bi_chart_interobserver} and \Cref{tab:interobserver_clinical_metrics}, as well as the detailed Bland-Altman plots shown in \Cref{fig:Bi_chart_interobserver}. The results, summarized in \Cref{tab:interobserver_segmentation} and \Cref{fig:pair-t-test}, reveal that the top-performing AI algorithms are approaching the level of human experts for segmenting the LV and MYO in images with mild to intermediate motion artifacts. For these structures, there was no statistically significant difference between the best AI model and the expert annotator.

However, a performance gap remains, particularly for the RV and in images with severe artifacts. Human experts consistently outperformed the AI models in RV segmentation across all quality levels and demonstrated greater robustness in segmenting all structures in severely degraded images. This suggests that while AI is nearing human-level performance under ideal conditions, improving robustness to severe artifacts and mastering the complex geometry of the RV are key areas for future work.

\begin{table*}[htbp]
  \centering
  \caption{Evaluation of interobserver agreement of segmentation masks. The reported data are in the format of mean $\pm$ standard deviation. $\uparrow$ represents higher is better,  $\downarrow$ represents lower is better.}
  \resizebox{0.8\linewidth}{!}{
    \begin{tabular}{cccccccc}
    \toprule
    \multirow{2}[4]{*}{IQA Score} & \multicolumn{3}{c}{DSC (\%) $\uparrow$} &       & \multicolumn{3}{c}{HD95 ($mm$)$\downarrow$} \\
    \cmidrule{2-4}\cmidrule{6-8}          & LV    & MYO   & RV    &       & LV    & MYO   & RV \\
    \midrule
    1     & 95.37 $\pm$ 2.02 & 88.06 $\pm$ 3.63 & 95.54 $\pm$ 2.64 &       & 1.72 $\pm$ 2.06 & 1.50 $\pm$ 0.56 & 1.59 $\pm$ 2.12 \\
    2     & 93.55 $\pm$ 2.16 & 86.05 $\pm$ 2.42 & 93.75 $\pm$ 3.26 &       & 2.98 $\pm$ 3.27 & 3.18 $\pm$ 3.19 & 2.26 $\pm$ 2.19 \\
    3     & 93.73 $\pm$ 0.95 & 86.76 $\pm$ 2.21 & 93.00 $\pm$ 2.33 &       & 1.55 $\pm$ 0.23 & 1.92 $\pm$ 0.06 & 2.41 $\pm$ 1.08 \\
    \bottomrule
    \end{tabular}%
  }
  \label{tab:interobserver_segmentation}%
\end{table*}%

\begin{figure*}[htbp]
  \centering
  \includegraphics[width=\linewidth]{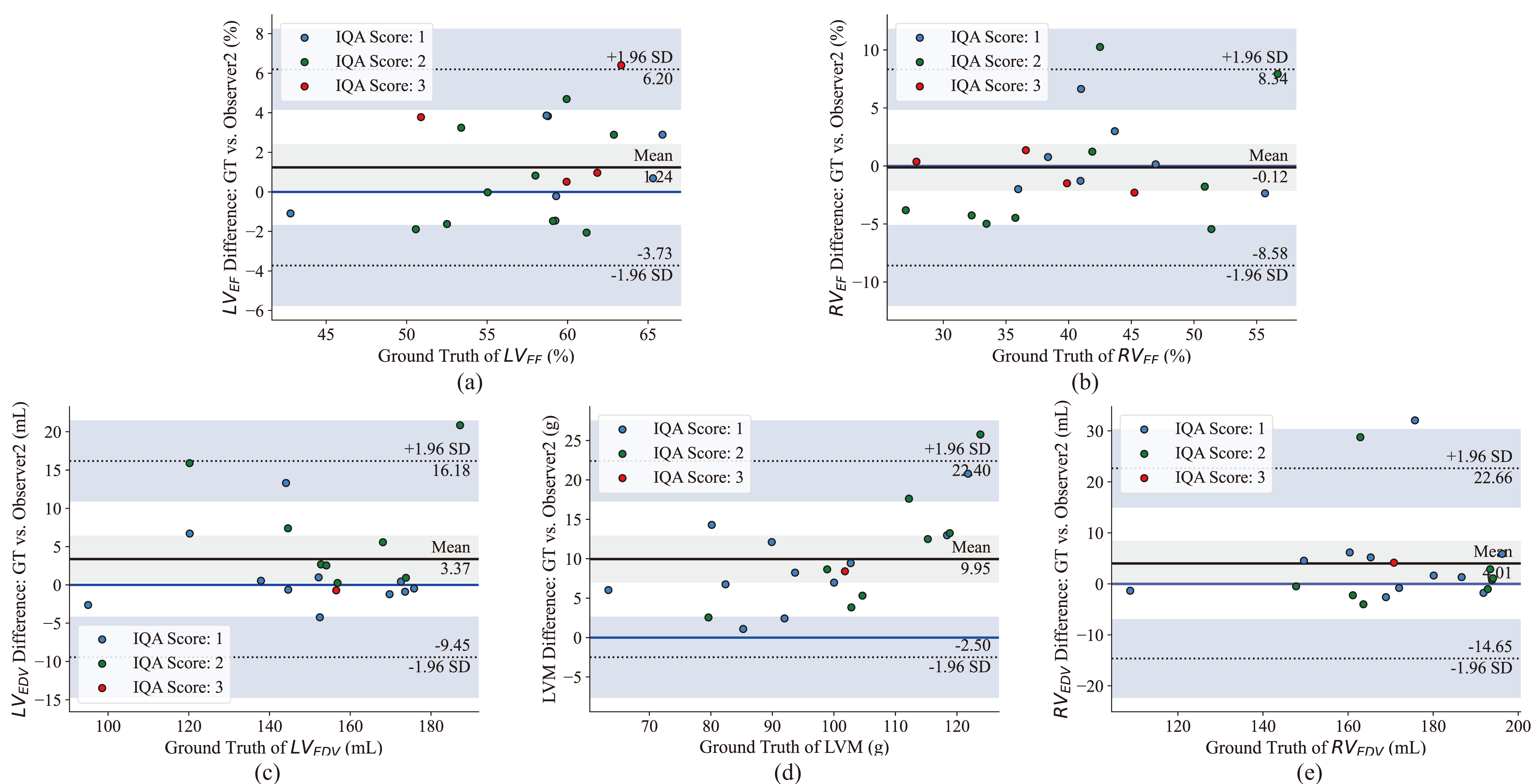}
  \caption{Bland-Altman plots illustrating the inter-observer agreement for the five clinical biomarkers. The solid black line indicates the mean difference, and the dashed lines represent the 95\% limits of agreement.}
  \label{fig:BA_plot_interobserver}
\end{figure*}

\begin{center}
  \centering
  \includegraphics[width=\linewidth]{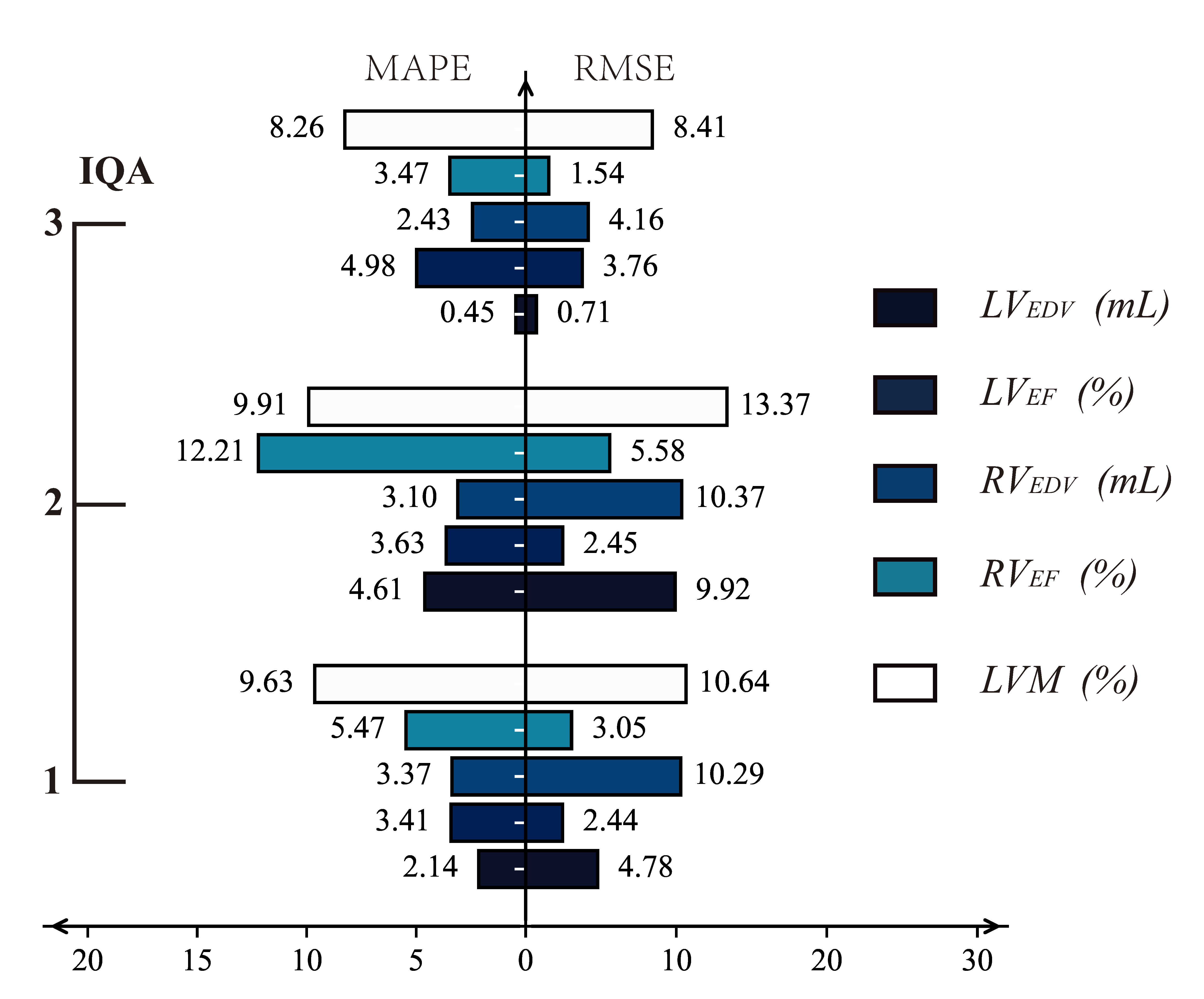}
  \captionof{figure}{Bidirectional bar chart of clinical indicator errors between two human observers, stratified by IQA score. The left bars show the MAPE, and the right bars show the RMSE.}
  \label{fig:Bi_chart_interobserver}
\end{center}

\begin{center}
  \centering
  \captionof{table}{Differences of clinical indicators between independent observers.}
  \label{tab:interobserver_clinical_metrics}%
  \resizebox{\linewidth}{!}{
    \begin{tabular}{lrrrrrr}
      \toprule
      \multicolumn{1}{c}{\multirow{2}[4]{*}{Metrics}} & \multicolumn{2}{c}{IQA 1} & \multicolumn{2}{c}{IQA 2} & \multicolumn{2}{c}{IQA 3} \\
  \cmidrule{2-7}          & \multicolumn{1}{c}{MAE} & \multicolumn{1}{c}{RMSE} & \multicolumn{1}{c}{MAE} & \multicolumn{1}{c}{RMSE} & \multicolumn{1}{c}{MAE} & \multicolumn{1}{c}{RMSE} \\
    \midrule
    $LV_{EDV}$ ($mL$) & 2.91  & 4.78  & 7.01  & 9.92  & 0.71  & 0.71 \\
    $LV_{EF}$ (\%) & 2.00  & 2.44  & 2.08  & 2.45  & 2.91  & 3.76 \\
    $RV_{EDV}$ ($mL$) & 5.76  & 10.29 & 5.16  & 10.37 & 4.16  & 4.16 \\
    $RV_{EF}$ (\%) & 2.31  & 3.05  & 4.91  & 5.58  & 1.38  & 1.54 \\
    $LVM$ ($g$) & 9.19  & 10.64 & 11.18 & 13.37 & 8.41  & 8.41 \\
    \bottomrule
    \end{tabular}%
  }
\end{center}%

\subsection{Limitations and future directions}
The \textit{CMRxMotion} challenge has provided a valuable benchmark, but it also highlights several limitations and opportunities for future research.

\textbf{Advancing model architectures:} The top-performing solutions in this challenge primarily relied on established architectures like nnU-Net and EfficientNet. Future work could explore the potential of newer architectures, such as Vision Foundation Models, which may offer enhanced feature representation capabilities. Furthermore, the exploration of more sophisticated multi-task and self-supervised learning frameworks could lead to improved generalization and robustness.

\textbf{Clinical translation and efficiency:} A significant barrier to the clinical adoption of many of the submitted algorithms is their computational cost. As shown in our efficiency analysis (\Cref{fig:efficiency}), the most accurate models were often the most computationally intensive. Future research should focus on model optimization and the development of lightweight architectures that can provide rapid inference without sacrificing accuracy, a prerequisite for real-time clinical applications.

\textbf{Synergy with image reconstruction:} This challenge focused on analyzing already-acquired images. A promising future direction is the synergistic development of segmentation algorithms and accelerated, motion-corrected MRI reconstruction techniques. By improving the quality of the input images at the source, the challenge of robust segmentation could be substantially mitigated.

\textbf{Dataset diversity:} The challenge dataset, while carefully curated, was based on simulated motion from healthy volunteers at a single center. To develop truly generalizable models, future efforts should focus on collecting large-scale, multi-center datasets that include real-world clinical data from diverse patient populations with a wide range of pathologies. This will be essential for training and validating models that are robust to the full spectrum of challenges encountered in routine clinical practice.

\section{Conclusion}
The \textit{CMRxMotion} challenge established the first public benchmark for evaluating the robustness of automated CMR analysis algorithms against respiratory motion artifacts. Through a unique dataset with a controlled spectrum of motion, we provided a standardized framework for assessing automated image quality assessment and robust cardiac segmentation.

The results from 22 submitted algorithms demonstrate that while automated IQA is feasible (top Kappa score of 0.631), accurately classifying the severity of motion remains an open challenge. In the RCS task, state-of-the-art methods achieved high accuracy on images with minimal motion. However, our analysis revealed a crucial finding: segmentation performance, and consequently the accuracy of derived clinical biomarkers, degrades significantly as motion artifacts increase. A notable performance gap also persists between the best algorithms and human experts, particularly in cases of severe motion. These results underscore the critical need for continued research into motion-robust analysis techniques, for which the \textit{CMRxMotion} dataset will serve as a valuable, ongoing resource.

\section*{Declaration of interests}
The authors declare that they have no known competing financial interests or personal relationships that could have appeared to influence the work reported in this paper.

\section*{Acknowledgments}
Shuo Wang was supported in part by the National Key Research and Development Program of China (No. 2024YFF1207500), Shanghai Municipal Education Commission Project for Promoting Research Paradigm Reform and Empowering Disciplinary Advancement through Artificial Intelligence (SOF101020), and the International Science and Technology Cooperation Program under the 2023 Shanghai Action Plan for Science (23410710400). Chengyan Wang is supported by the Shanghai Municipal Science and Technology Major Project (No.2023SHZDZX02A05), the Shanghai Rising-Star Program (No.24QA2703300), and the National Key R\&D Program of China (2024YFC3405800). Kang Wang is supported by Opening Laboratory Program of Linyi People's Hospital (No.2024LYKC001). Shi Zhang is supported by the National Natural Science Foundation of China (No. 82202145) and the Shanghai Magnolia Talent Program Pujiang Project (No. 2023PJD012).

We would like to thank all volunteers who generously took the time to participate in this study. We also extend our gratitude to Synapse and Paratera Technology Ltd for their technical support. The computations were performed using the CFFF platform of Fudan University.

\section*{CRediT authorship contribution statement}
\textbf{Kang Wang}: Methodology, Software, Validation, Formal analysis, Investigation, Data Curation, Writing - Original Draft, Writing - Editing, Visualization, Funding acquisition; \textbf{Chen Qin}: Project administration, Conceptualization, Resources, Writing - Review \& Editing, Supervision; \textbf{Zhang Shi}: Validation, Resources, Data curation, Funding acquisition; \textbf{Haoran Wang}: Software, Validation, Formal analysis, Data Curation; \textbf{Xiwen Zhang}:Formal analysis, Writing - Review \& Editing, Data Curation; \textbf{Chen Chen}: Formal analysis, Validation, Resources, Writing - Review \& Editing; \textbf{Cheng Ouyang}: Formal analysis, Validation, Resources, Writing - Review \& Editing; \textbf{Chengliang Dai}: Validation, Resources, Data curation; \textbf{Yuanhan Mo}: Validation, Resources, Data curation; \textbf{Chenchen Dai}: Validation, Resources, Data curation; \textbf{Xutong Kuang}: Validation, Resources, Data curation; \textbf{Ruizhe Li}: Writing - Review \& Editing, Methodology, Software, Formal analysis; \textbf{Xin Chen}: Writing - Review \& Editing, Methodology, Software, Formal analysis; \textbf{Xiuzheng Yue}: Writing - Review \& Editing, Methodology, Software, Formal analysis; \textbf{Song Tian}: Writing - Review \& Editing, Methodology, Software, Formal analysis; \textbf{Alejandro Mora-Rubio}: Writing - Review \& Editing, Methodology, Software, Formal analysis; \textbf{Kumaradevan Punithakumar}: Writing - Review \& Editing, Methodology, Software, Formal analysis; \textbf{Shizhan Gong}: Writing - Review \& Editing, Methodology, Software, Formal analysis; \textbf{Qi Dou}: Writing - Review \& Editing, Methodology, Software, Formal analysis; \textbf{Sina Amirrajab}: Writing - Review \& Editing, Methodology, Software, Formal analysis; \textbf{Yasmina Al Khalil}: Writing - Review \& Editing, Methodology, Software, Formal analysis; \textbf{Cian M. Scannell}: Writing - Review \& Editing, Methodology, Software, Formal analysis; \textbf{Lexiaozi Fan}: Writing - Review \& Editing, Methodology, Software, Formal analysis; \textbf{Huili Yang}: Writing - Review \& Editing, Methodology, Software, Formal analysis; \textbf{Xiaowu Sun}: Writing - Review \& Editing, Methodology, Software, Formal analysis; \textbf{Rob van der Geest}: Writing - Review \& Editing, Methodology, Software, Formal analysis; \textbf{Tewodros Weldebirhan Arega}: Writing - Review \& Editing, Methodology, Software, Formal analysis; \textbf{Fabrice Meriaudeau}: Writing - Review \& Editing, Methodology, Software, Formal analysis; \textbf{Caner Özer}: Writing - Review \& Editing, Methodology, Software, Formal analysis; \textbf{Amin Ranem}: Writing - Review \& Editing, Methodology, Software, Formal analysis; \textbf{John Kalkhof}: Writing - Review \& Editing, Methodology, Software, Formal analysis; \textbf{İlkay Öksüz}: Writing - Review \& Editing, Methodology, Software, Formal analysis; \textbf{Anirban Mukhopadhyay}: Writing - Review \& Editing, Methodology, Software, Formal analysis; \textbf{Abdul Qayyum}: Writing - Review \& Editing, Methodology, Software, Formal analysis; \textbf{Moona Mazher}: Writing - Review \& Editing, Methodology, Software, Formal analysis; \textbf{Steven A Niederer}: Writing - Review \& Editing, Methodology, Software, Formal analysis; \textbf{Carles Garcia-Cabrera}: Writing - Review \& Editing, Methodology, Software, Formal analysis; \textbf{Eric Arazo}: Writing - Review \& Editing, Methodology, Software, Formal analysis; \textbf{Michal K. Grzeszczyk}: Writing - Review \& Editing, Methodology, Software, Formal analysis; \textbf{Szymon Płotka}: Writing - Review \& Editing, Methodology, Software, Formal analysis; \textbf{Wanqin Ma}: Writing - Review \& Editing, Methodology, Software, Formal analysis; \textbf{Xiaomeng Li}: Writing - Review \& Editing, Methodology, Software, Formal analysis; \textbf{Rongjun Ge}: Writing - Review \& Editing, Methodology, Software, Formal analysis; \textbf{Yongqing Kou}: Writing - Review \& Editing, Methodology, Software, Formal analysis; \textbf{Xinrong Chen}: Validation, Resources, Data curation; \textbf{He Wang}:Formal analysis, Resources, Writing - Review \& Editing; \textbf{Chengyan Wang}: Conceptualization, Resources, Writing - Review \& Editing, Funding acquisition; \textbf{Wenjia Bai}: Conceptualization, Supervision, Formal analysis, Resources, Writing - Review \& Editing, Software; \textbf{Shuo Wang}: Project administration, Conceptualization, Methodology, Validation, Formal analysis, Data curation, Writing - Original Draft, Writing - Review \& Editing, Supervision, Funding acquisition.

\section*{Data availability}
Access to the challenge data will be granted upon approval.

\bibliographystyle{model2-names.bst}\biboptions{authoryear}
\bibliography{refs}

\end{multicols}

\end{document}